\documentclass[10pt,reqno]{article}
\usepackage[b5paper,top=1.2in, left=0.9in]{geometry}
\usepackage{verbatim}
\usepackage{amssymb}
\usepackage{amsmath}
\usepackage{graphicx}
\usepackage{appendix}
\usepackage{color}
\usepackage{amsthm}
\usepackage{tikz}
\usepackage{ifthen}
\usepackage{float}
\usepackage{subfloat}
\usepackage{caption}
\usepackage{subcaption}
\usetikzlibrary{arrows,positioning,decorations.pathmorphing,  decorations.markings}

\usepackage{cite} 
\usepackage[colorlinks, allcolors=blue!70!black, linktocpage]{hyperref}
\usepackage[inline]{enumitem}
\usepackage[utf8]{inputenc}

\usepackage{epic}
\usepackage{eepic}
\usepackage{graphicx}

\numberwithin{equation}{section}

\usepackage{bm}

\DeclareMathOperator*{\SumInt}{%
\mathchoice%
  {\ooalign{$\displaystyle\sum$\cr\hidewidth$\displaystyle\int$\hidewidth\cr}}
  {\ooalign{\raisebox{.14\height}{\scalebox{.7}{$\textstyle\sum$}}\cr\hidewidth$\textstyle\int$\hidewidth\cr}}
  {\ooalign{\raisebox{.2\height}{\scalebox{.6}{$\scriptstyle\sum$}}\cr$\scriptstyle\int$\cr}}
  {\ooalign{\raisebox{.2\height}{\scalebox{.6}{$\scriptstyle\sum$}}\cr$\scriptstyle\int$\cr}}
}

\begin{document}
\begin{titlepage}

 \vskip 2.7 cm

\centerline{\Large \bf Integrable lattice spin models from supersymmetric dualities}
\vskip 1.5 cm

\centerline{\large {\bf Ilmar Gahramanov\footnote{ilmar.gahramanov@msgsu.edu.tr}$\,^{a,b,c}$} and
{\bf Shahriyar Jafarzade\footnote{shahriyar.jzade@gmail.com}$\,^{a,b}$} }

\begin{center}
\textit{$^{a}$ Department of Physics, Mimar Sinan Fine Arts University,\\ Bomonti 34380, Istanbul, Turkey} \\
\vspace{2.8mm}
\textit{$^{b}$ Department of Mathematics, Khazar University, \\ Mehseti St. 41, AZ1096, Baku, Azerbaijan} \\
\vspace{2.8mm}
\textit{$^{c}$  Max Planck Institute for Gravitational Physics (Albert Einstein Institute),\\ Am M\"{u}hlenberg 1, D-14476 Potsdam, Germany} \\
\texttt{} \\
\vspace{0mm}
\end{center}

\vskip 0.5cm
\hfill {\textbf{Dedicated to the memory of Ludvig D. Faddeev}}

\vskip 1.5cm \centerline{\bf Abstract} \vskip 0.2cm \noindent
Recently, there has been observed an interesting correspondence between supersymmetric quiver gauge theories with four supercharges and integrable lattice models of statistical mechanics such that the two-dimensional spin lattice is the quiver diagram, the partition function of the lattice model is the partition function of the gauge theory and the Yang-Baxter equation expresses the identity of partition functions for dual pairs.  This correspondence is a powerful tool which enables us to generate new integrable models. The aim of the present paper is to give a short account on a progress in integrable lattice models which has been made due to the relationship with supersymmetric gauge theories.

\vskip 1.1cm

{\small  {\textbf {MSC classes:} 81T60, 16T25, 14K25, 33D60, 33E20, 33D90, 39A13 }}


\end{titlepage}

\small
\tableofcontents
\normalsize
\setcounter{page}{1}
\section{Introduction}

Integrability is a beautiful phenomenon which plays an important role in theoretical and mathematical physics. One of the key structural elements leading to quantum integrability is the Yang-Baxter equation \cite{McGuire:1964zt,Yang:1967bm,Baxter:1972hz,Baxter:1982zz,Jimbo:1989qm,Kulish:1980ii}
\begin{equation}
\mathbb{R}_{12} (u-v)\,\mathbb{R}_{13}(u)\, \mathbb{R}_{23}(v)
=\mathbb{R}_{23}(v)\,\mathbb{R}_{13}(u)\,\mathbb{R}_{12}(u-v) \;,
\end{equation}
where the operators $\mathbb{R}_{ik}(u)$ act in the tensor product of some three vector spaces $\mathbb{V} \otimes \mathbb{V} \otimes \mathbb{V}$ and depend on the spectral parameter $u$. The importance of the Yang-Baxter equation as a condition for integrability was noticed by Ludvig Faddeev who developed (with his Leningrad group) a deep connection between integrability and other areas of mathematical physics.  Nowadays the Yang-Baxter equation has a relation to quantum field theory, knot theory, string theory, statistical physics, conformal field theory etc. 

In this work we consider quantum integrability of two-dimensional square lattice spin models of statistical mechanics with the pair interaction between neighboring spins. The most known example of such models is the two-dimensional Ising model \cite{Ising:1925em} which was solved by Onsager \cite{Onsager:1943jn}. Onsager also observed that the Boltzmann weights of the Ising model satisfies the star-triangle relation which is a special form \cite{Baxter:1997tn} of the Yang-Baxter equation for integrable statistical models with spin variables living on sites of the lattice:
\begin{align}\nonumber
\sum_{\sigma _0}\mathcal{S}( \sigma _0)\mathcal{W}_{ \eta - \alpha}( \sigma _i , \sigma _0)&\mathcal{W}_{ \eta - \beta}( \sigma _j, \sigma _0) \mathcal{W}_{ \eta - \gamma}( \sigma _k, \sigma _0) \\
 =\, & \mathcal{R}(\alpha,\beta,\gamma)\mathcal{W}_{  \alpha}( \sigma _j, \sigma _k)W_{ \beta}( \sigma _i, \sigma _k) \mathcal{W}_{  \gamma}( \sigma _j, \sigma _i) \;, 
\end{align}
where $\mathcal{W}$ and $S$ stand for the Boltzmann weight functions of the model. The star-triangle relation appears as a condition for commuting transfer matrices what makes the model integrable \cite{Baxter:1982zz,au1989onsager}.

There have been many developments in the integrable lattice spin models since Onsager's solution. There are by now many solutions\footnote{Classification of all solutions \cite{Kulish:1980ii,Kulish:1981gi} to the quantum Yang-Baxter equation still remains an open problem.} of the star-triangle equation, i.e. Ising-like lattice models, most notable ones are the Fateev-Zamodchikov model \cite{Fateev:1982wi} (the case N=2 gives the Ising model), Kashiwara-Miwa model \cite{Kashiwara:1986tu, Hasegawa:1990du, Gaudin:1990gf}, chiral Potts model \cite{vonGehlen:1984bi, AuYang:1987zc, Baxter:1987eq}, Faddeev-Volkov model \cite{Faddeev:1994fw,Volkov:1992uv}, Bazhanov-Sergeev model \cite{Bazhanov:2010kz} etc.

One of the most surprising developments in the field has appeared recently \cite{Spiridonov:2010em} coming from a different area of theoretical physics. It was observed a relationship between exact results in supersymmetric quiver gauge theories and exactly solvable two-dimensional lattice models in statistical mechanics \cite{Yamazaki:2012cp,Yamazaki:2013nra,Yagi:2015lha,Yamazaki:2015voa,Gahramanov:2015cva,Kels:2015bda,Maruyoshi:2016caf,Gahramanov:2016ilb,Yamazaki:2016wnu,Yagi:2016oum,Yagi:2017hmj,Kels:2017toi,Kels:2017fyt}.  In the  gauge/YBE correspondence, as it is called\footnote{The name ``gauge/YBE'' was first used by Yamazaki in \cite{Yamazaki:2012cp} probably in an analogue to the gauge/Bethe correspondence \cite{Nekrasov:2009rc,Nekrasov:2009uh}.}, the integrability in statistical models is a direct consequence of supersymmetric duality. Roughly speaking, gauge/YBE correspondence relates the Yang-Baxter equation with the equality of partition functions for supersymmetric dual theories. This relationship has led to the construction of new exactly solvable models of statistical mechanics and we believe that much more are to be found.

In this work we try to present an elementary description of the gauge/YBE correspondence and to list solutions of the star-triangle relation found (or related) by this correspondence. Of course, it is impossible to give all details of this recent subject of research, therefore in some places these notes have a sketchy character. We hope to convince the reader, both with mainly integrability background and supersymmetry alike, that the subject has many interesting applications and new open problems.

\vspace{0.3cm}

The rest of the paper is organized in the following way:

\begin{itemize}

\item In Section 2 we review integrable lattice models and formulate the star-triangle relation (Yang-Baxter equation) for Ising-like lattice models of statistical mechanics. 

\item In Section 3 we give a very brief review of supersymmetric duality, exact results for partition functions and quiver notations.

\item We present a survey of recent progress in integrable statistical models inspired by supersymmetric gauge theory computations in Section 4 and list all recently found solutions to the star-triangle relation. 

\item The paper concludes with comments on the recent status of the correspondence and briefly discusses some open problems in Section 5. 
\end{itemize}

\section{A crash course on integrable lattice models}

The main players in the notes are solvable\footnote{The terms ``solvable'' and ``integrable'' are the same in the context of this paper and will be used interchangeably.} lattice models and quiver gauge theories with four supercharges in two, three and four dimensions. In this section, we set up basic terminology about the exactly solvable lattice models of statistical mechanics. More details on the subject can be found in the book by Baxter \cite{Baxter:1982zz} and in the review papers \cite{Bazhanov:2016ajm,Baxter:1997tn,perk2006yang,Bellon:1992sf,deguchi2003statistical,Saleur:1990uz}. The section will mainly follow the exposition in \cite{Bazhanov:2016ajm,Gahramanov:2016ilb,Kels:2017fyt}.

\subsection{Lattice models in statistical mechanics}

An Ising-like model on a two-dimensional square lattice is defined as follows. At each site $i$ there is a ``spin'' variable $\sigma_i$ which takes some set of continuous or discrete values (or both as we will see later) in some range. Two adjacent spins $i$ and $j$ interact with an energy $\epsilon(\sigma_i, \sigma_j)$. The quantities of interest in statistical physics are statistical sums, such as the following partition function
\begin{equation}
Z= \sum_{\{\sigma\}} e^{-\frac{E(\sigma)}{k_BT}}
\end{equation}
where the summation runs over all values of spins;  $E(\sigma), k_{B}$ and $T$   are the energy of the system,  Boltzmann constant and  temperature  respectively.
Let
\begin{equation}
\mathcal{W}(\sigma_i, \sigma_j) \ = \ e^{-\frac{\epsilon(\sigma_i, \sigma_j)}{k_B T}}
\end{equation}
be the Boltzmann weight of the edge $(i,j)$, with an interaction energy $\epsilon(\sigma_i,\sigma_j)$ between spins $\sigma_i$ and $\sigma_j$. Then the partition function can be written in the following way
\begin{equation} \label{pf}
Z \ = \ \sum \prod_{<i,j>}\mathcal{W}(\sigma_i, \sigma_j)
\end{equation}
The ``integrable model'' means that one can evaluate the partition function (\ref{pf}) in the thermodinamic limit $N \rightarrow \infty$, where $N$ is a number of sites of the lattice.

There exist  two other types of models in two-dimensional statistical mechanics, the IRF model and the vertex model\footnote{As was mentioned above we will not be interested in the solutions of IRF and vertex-type Yang-Baxter equation and talk about them here only for completeness. We would like to point out that for some models in our list discussed in the next sections the IRF and vertex-type solutions are known. We will not discuss those solution and refer to the original papers \cite{Yamazaki:2015voa,Gahramanov:2015cva,Jafarzade:2017fsc,Derkachov:2010zz,Derkachov:2012iv,Chicherin:2014dya}.}.

In the ``interaction round a face model'' (IRF) version of spin models four spins round a face of the lattice interact with each other. This interaction can be determined by the energy of face $\varepsilon(\sigma_1, \sigma_2, \sigma_3, \sigma_4)$ depending on four spins. 
The notable examples of IRF models are the hard hexagon model \cite{baxter1980hard}, the cyclic solid-on-solid model \cite{Baxter:1972wg,Baxter:1972wf,Baxter:1972wh,Pearce:1988en,Pearce:1989ek} and the restricted solid-on-solid model \cite{Andrews:1984af}.

In the vertex model, spin variables are located on the edges of the spin lattice. In this case one associates the local Boltzmann weight with each vertex configuration, namely statistical weights depend on four spins surrounding each site. The most known examples of vertex models are the six-vertex model \cite{Lieb:1967zz,sutherland1967exact}, the eight-vertex model \cite{Baxter:1972hz} and the nineteen-vertex model \cite{Izergin:1980pe}. 

\begin{figure}[h]
    \centering
    \includegraphics[width=0.91\textwidth]{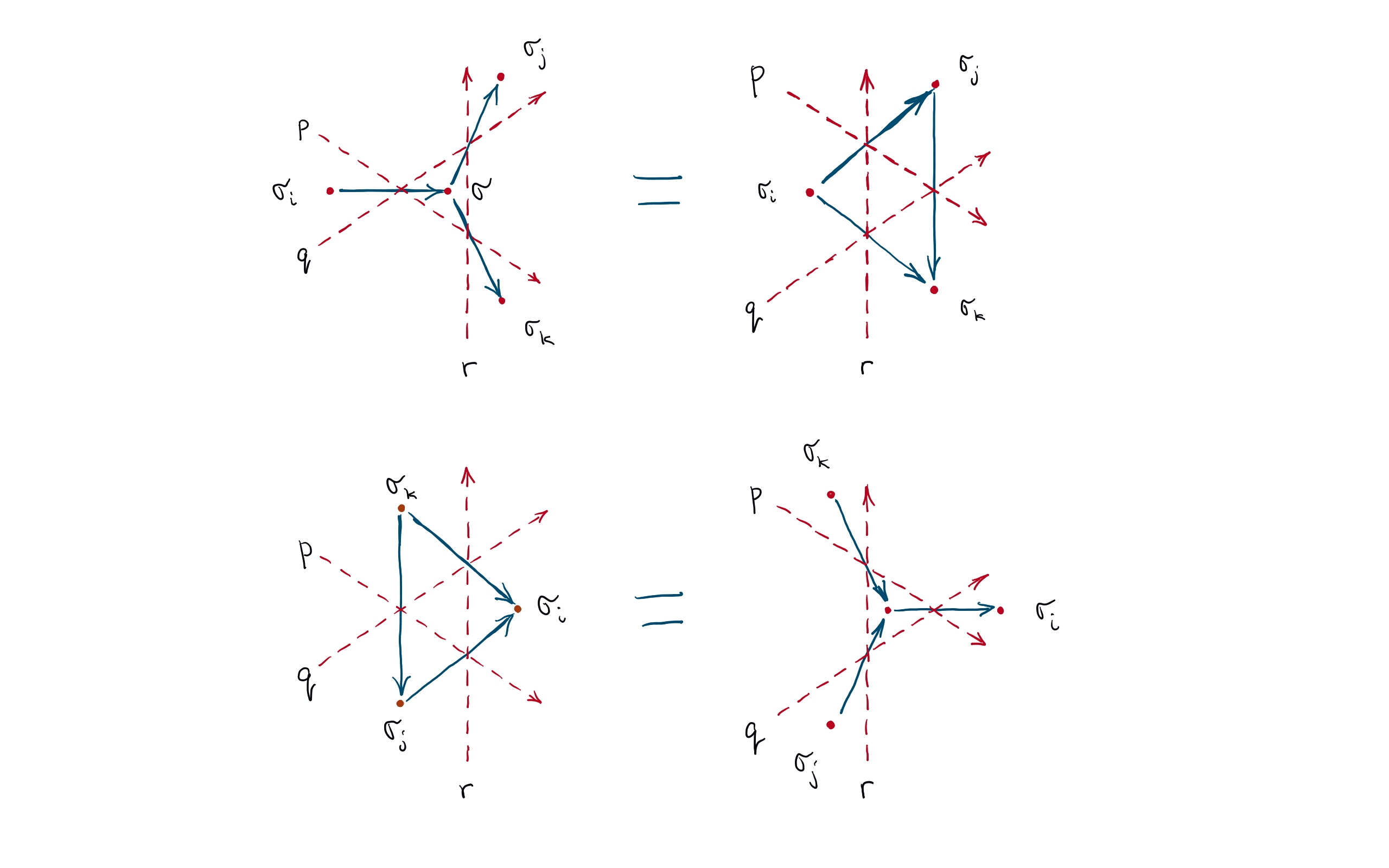}
   \caption{Star-triangle relation}
\end{figure}

\subsection{Integrability of models: star-triangle relation}

The search for the Ising-like integrable models can be reduced to the problem of finding the Boltzmann weights that satisfy the so-called star-triangle relation\footnote{Note that in the literature ``the star-triangle relation'' often is used also for the IRF-type models, see, e.g. \cite{Perk:1986nr}.}

\begin{align} \nonumber
& \sum_\sigma \mathcal{S}(\sigma)  \overline{\mathcal{W}}_{qr}(\sigma,\sigma_j)\mathcal{W}_{pr}(\sigma,\sigma_k) \overline{\mathcal{W}}_{pq}(\sigma_i,\sigma) \\
& \qquad \qquad \qquad \qquad \qquad = \mathcal{R}(p,q,r)\mathcal{W}_{pq}(\sigma_j,\sigma_k) \overline{\mathcal{W}}_{pr}(\sigma_i,\sigma_j)\mathcal{W}_{qr}(\sigma_i,\sigma_k), \\ \nonumber
& \sum_\sigma \mathcal{S}(\sigma)  \overline{\mathcal{W}}_{pq}(\sigma,\sigma_i)\mathcal{W}_{pr}(\sigma_k,\sigma) \overline{\mathcal{W}}_{qr}(\sigma_j,\sigma) \\
& \qquad \qquad \qquad \qquad \qquad = \mathcal{R}(p,q,r)\mathcal{W}_{pq}(\sigma_k,\sigma_j) \overline{\mathcal{W}}_{pr}(\sigma_j,\sigma_i)\mathcal{W}_{qr}(\sigma_k,\sigma_i).
\end{align}
where 
\begin{itemize}
\item Summation is over all spin variables, $\mathcal{W}_{pq}(\sigma_i,\sigma_j) $ and $\overline{\mathcal{W}}_{pq}(\sigma_i,\sigma_j) $  are two different kinds of  the Boltzmann weights describe the interaction between two spins;
\begin{figure}[h]
    \centering
    \includegraphics[width=0.45\textwidth]{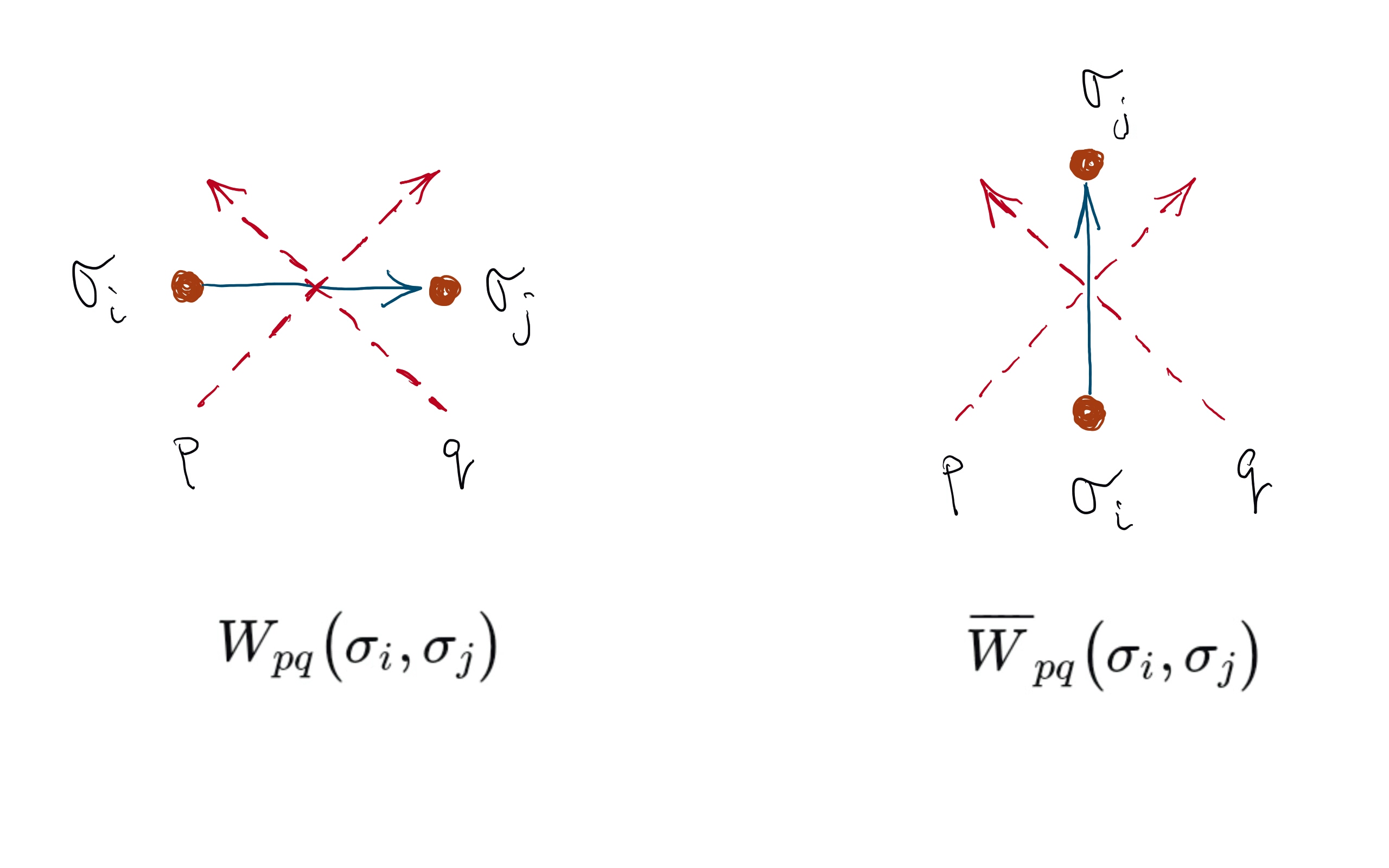}
       \caption{Two types of the Boltzmann weights}
\end{figure}

\item $\mathcal{S}(\sigma)$ is the rapidity-independent single-spin Boltzmann weight  assigned for each spin $\sigma$ on the lattice;
\item $ \mathcal{R}(p,q,r)$ is some factor depending on three rapidity variables and independent of the spins. 
\end{itemize}
The star-triangle relation is a sufficient condition for the existence of  an infinite set of commuting transfer matrices\footnote{Here we do not discuss a transfer matrix method and relation of the star-triangle equation to commutativity of transfer matrices, the interested reader can be find details in many places, for instance, in \cite{Baxter:1982zz,deguchi2003statistical}.} and thereby the model can be exactly solved using the transfer matrix method \cite{Baxter:1982xp}. Namely by defining the transfer matrices
\begin{align}
(T_q)_{\sigma, \bar{\sigma}} & = \prod_{i=1}^{L} \mathcal{W}_{pq}(\sigma_i, \bar{\sigma}_i) \overline{\mathcal{W}}_{pq}(\sigma_{i+1}, \bar{\sigma}_i)\\
(\bar{T}_r)_{\sigma, \bar{\sigma}} & = \prod_{i=1} ^{L}\overline{\mathcal{W}}_{pr}(\sigma_i, \bar{\sigma}_i) {\mathcal{W}}_{pr}(\sigma_{i}, \bar{\sigma}_{i+1})
\end{align}
with periodic boundary conditions that $\sigma_{L+1}=\sigma_1$ and $\bar{\sigma}_{L+1}=\bar{\sigma}_1$ one can prove that $T_q \bar{T}_r=T_r \bar{T}_q$. If one has such a family of commuting transfer matrices then a partition function (\ref{pf}) can be calculated exactly .

In all our examples here, the two types of Boltzmann weights $\mathcal{W}_{pq}$, $\overline{\mathcal{W}}_{pq}$, depend on rapidity variables only via their difference\footnote{The majority of lattice models of statistical mechanics satisfy this property, the most notable exception being the Chiral Potts model \cite{Baxter:1987eq}.} $p-q$.  Consequently the Boltzmann weights will be written in terms of the spectral variable $\alpha=p-q$, as 
\begin{equation}
\mathcal{W}_\alpha(\sigma_i,\sigma_j):=\mathcal{W}_{pq}(\sigma_i,\sigma_j)\,, \quad \text{and} \quad\, \overline{\mathcal{W}}_\alpha(\sigma_i,\sigma_j):=\overline{\mathcal{W}}_{pq}(\sigma_i,\sigma_j) \,.
\end{equation}
The two Boltzmann weights are also related by the crossing symmetry 
\begin{equation}
\overline{\mathcal{W}}_\alpha(\sigma_i,\sigma_j)={\mathcal{W}}_{\eta-\alpha}(\sigma_i,\sigma_j) \,,
\end{equation}
where $\eta>0$ is a real valued, model dependent ``crossing parameter''.  Thus all two-spin interactions in the lattice model may be described in terms of the single Boltzmann weight ${\mathcal{W}}_\alpha(\sigma_i,\sigma_j)$. 

The Boltzmann weights considered here are spin reflection symmetric, i.e. unchanged by interchanging the spin variables $\sigma_i$ and $\sigma_j$:
\begin{equation}
\mathcal{W}_\alpha(\sigma_i,\sigma_j)=\mathcal{W}_\alpha(\sigma_j,\sigma_i) \;.
\end{equation}

The simple consequence of the star-triangle relation and initial condition gives the unitarity and inversion relations\footnote{Note that the inversion relation for the partition function may exist even for a model which is not integrable. }
\begin{align}\label{In-rel1}
\mathcal{W}_\alpha(\sigma_i,\sigma_j)\mathcal{W}_{-\alpha}(\sigma_i,\sigma_j)& =1
\end{align}
\begin{align}\label{In-rel2}\nonumber
\SumInt_{\sigma_0}\,S(\sigma_0) \mathcal{W}_{\eta-\alpha}(\sigma_i,\sigma_0) &\mathcal{W}_{\eta+\alpha}(\sigma_0,\sigma_j)  \\ &=\frac{1}{\mathcal{S}(\sigma_i)}(\delta(x_i\!+\!x_j)\,\delta_{m_i,-m_j}+\delta(x_i\!-\!x_j)\,\delta_{m_i,m_j})\,.
\end{align}

\begin{figure}[h]
    \centering
    \includegraphics[width=0.5\textwidth]{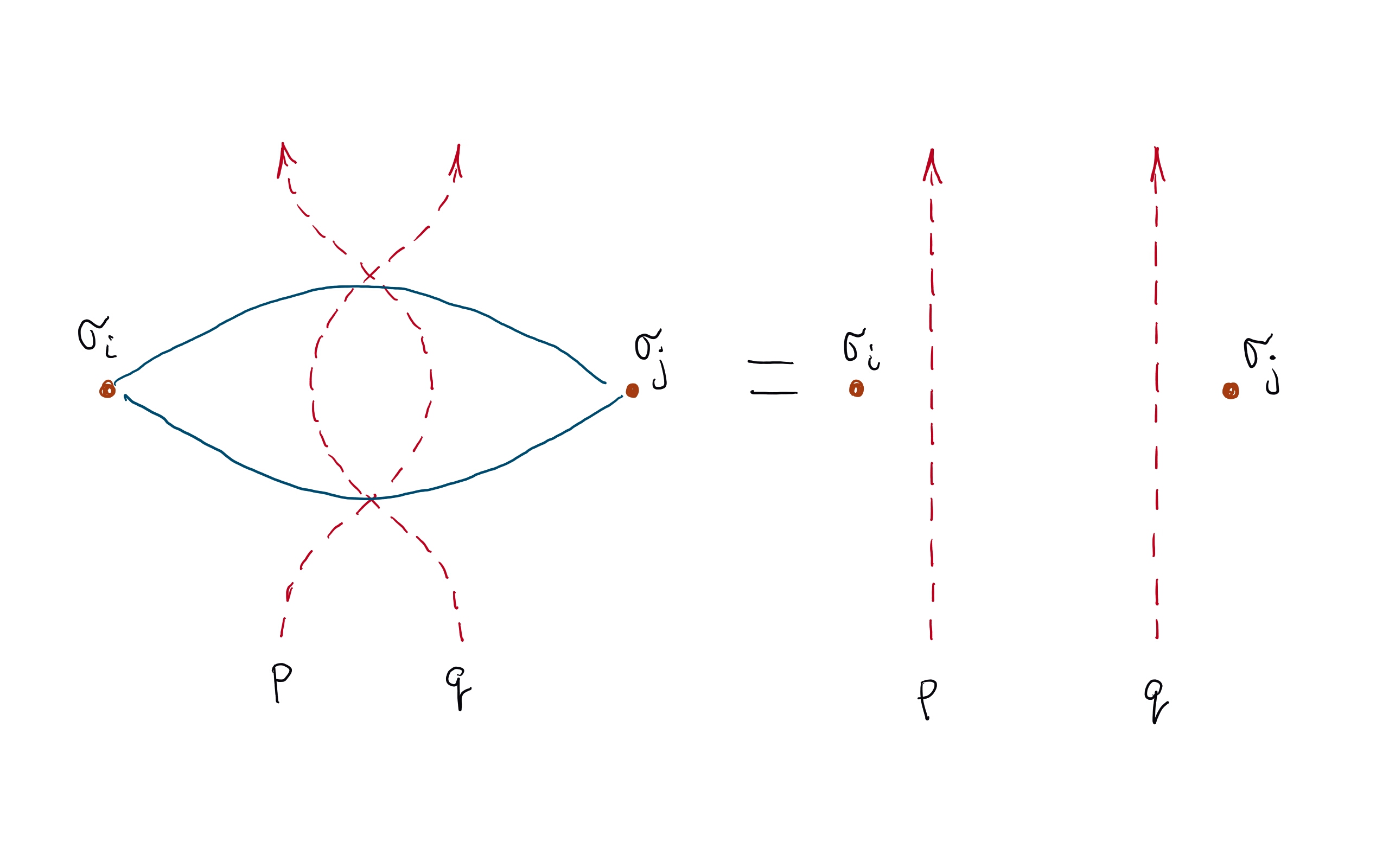}
   \caption{Inversion relation (\ref{In-rel1})}
\end{figure}
\begin{figure}[h]
    \centering
    \includegraphics[width=0.5\textwidth]{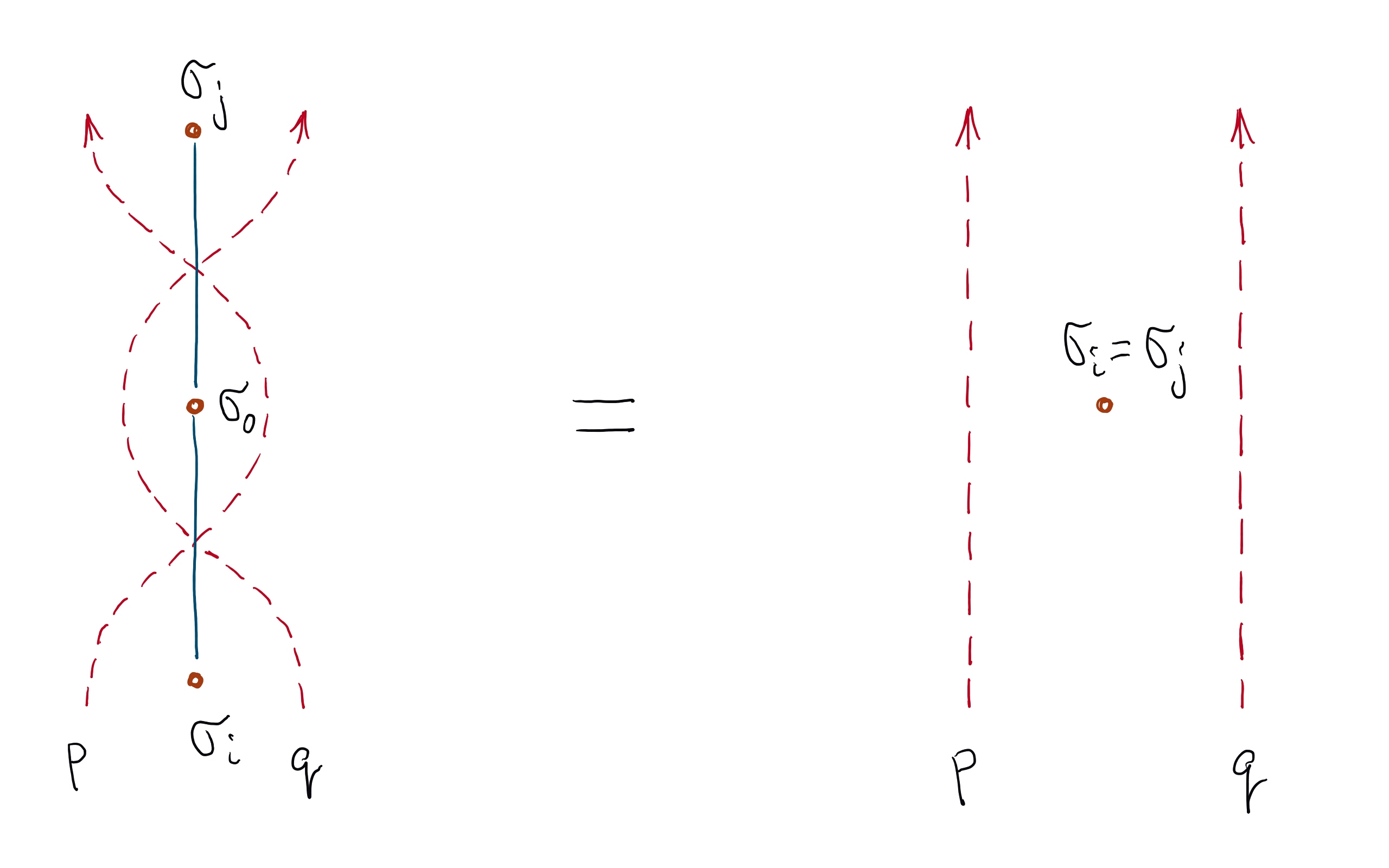}
   \caption{Inversion relation (\ref{In-rel2})}
\end{figure}



\section{A crash course on supersymmetric dualities and exact results}

Obviously, it is impossible to review supersymmetric dualities and exact results in supersymmetric gauge theories in a few pages. Our intention is to give some a short description of the important keywords on the subject.

\subsection{Supersymmetric duality}

In this section we very briefly remind some facts about supersymmetric duality. For more details, see e.g. \cite{Intriligator:1995au,Strassler:2001ue,Strassler:2005qs,Terning:2006bq}.

About two decades ago Seiberg \cite{Seiberg:1994pq} and many others found a non-trivial quantum equivalence between different supersymmetric theories, called supersymmetric duality. To be more precise it was shown that two or more different theories may describe the same physics in the far infrared limit, i.e. an observer testing the low energy physics (or physics at long distances) cannot distinguish the dual theories\footnote{It is worth mentioning that supersymmetric dual theories are not identical, but they give rise to the same physics at long distances.}. 

The supersymmetric duality was first constructed for four-dimensional ${\mathcal N}=1$ gauge theory with a matter in the fundamental representation. Later many examples of dualities have been found with complicated matter content, different gauge and flavor groups in different dimensions. Today, supersymmetric duality has become a key tool for studying strongly coupled effects.

The basic example of supersymmetric duality \cite{Seiberg:1994pq} is an $SU(N_c)$ electric gauge theory with $N_F$ flavors of quarks which possesses a dual description in terms of $N_f$ magnetic flavors of quarks charged under $SU(N_F-N_c)$ gauge group\footnote{In this case the gauge singlets of the dual theory interact with the flavors via the superpotential term.} in the so-called conformal window $\frac32 N_c < N_F < 3 N_c$. The field content of dual theories is summarized in the table below. These two theories flow to the same infrared fixed point.

\begin{center}
\begin{tabular}{|c||c|cc|}
\hline  
  & $SU(N_c)$ & $SU(N_F)_L$ & $SU(N_F)_R$ \\ \hline  
Q & $f$ & $f$ & 1 \\ 
Q' & $\bar{f}$ & 1 & $\bar{f}$ \\ 
\hline
\end{tabular}\\
Matter content of the electric theory.
\end{center}

\begin{center}
\begin{tabular}{|c||c|cc|}
\hline
  & $SU(N_F-N_c)$ & $SU(N_F)_L$ & $SU(N_F)_R$ \\ \hline
$q$ & $f$ & $\bar{f}$ & 1 \\ 
$q'$ & $\bar{f}$ & 1 & $f$  \\ 
$M$ & 1 & f & $\bar{f}$ \\ \hline
\end{tabular}\\
 Matter content of the magnetic theory.
\end{center}

The main point for us about the supersymmetric dualities is that the partition functions of dual theories are expected to be equal. In the context of gauge/YBE correspondence integrability on the statistical models' side is equivalent to the equality of partition functions of supersymmetric dual theories. It means that one may generate solutions to the Yang-Baxter equation by considering partition functions of suitable duality.


\subsection{Quiver gauge theories}

Here we briefly outline quiver notation of gauge theories which is a very useful tool for summarizing the group-theoretical data about a gauge theory in a compact way. Quiver gauge theories have been studied in physics more than forty years, initially, they were used in composite model building in the context of the Standard Model. For more details, see e.g. \cite{Berenstein:2002fi,Yamazaki:2008bt,Hanany:2005ss}.


Supersymmetric gauge theories considered in the work are specified with gauge group $G$ (in our examples we consider just $SU(2)$ group) and the matter fields transforming as chiral multiplets in a suitable representation. One can encode this information\footnote{Note that the superpotential term of the theory is not encoded by the quiver diagram.} in quiver diagrams using nodes for the gauge groups and edges for the matter multiplets.

Consider a theory with the gauge group $G$ as a direct product of simple groups $G_i$
\begin{equation}
G=G_1\times G_2 \times \ldots \times G_n \;.
\end{equation}
In the quiver diagram
\begin{itemize}
\item each node $g_i$ corresponds to a vector multiplet in the adjoint representation of a gauge group $G_i$;

\item each edge corresponds to the matter multiplet in the bifundamental representation. In general, one uses arrows between nodes. The arrow going from $g_i$ to $g_j$ corresponds to a chiral multiplet in the fundamental representation of $g_i$ and the anti-fundamental representation of $g_j$.
\end{itemize}

\begin{figure}[h] 
    \centering
    \includegraphics[width=0.97\textwidth]{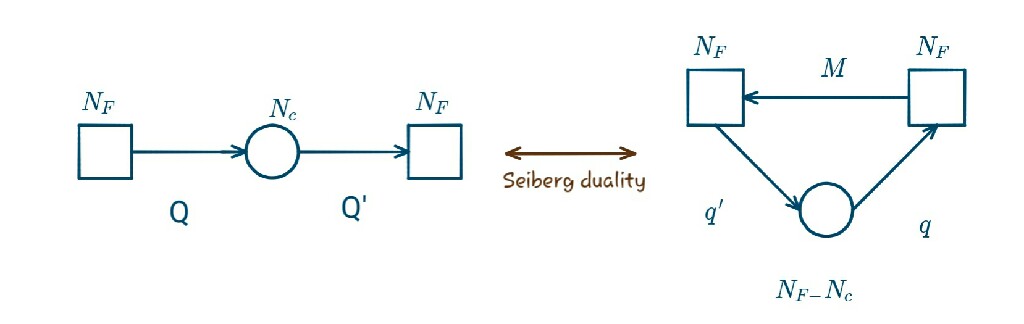}
   \caption{Quiver diagram for the Seiberg duality}
\label{seibergfig}
\end{figure}

The quiver diagram encoding the $SU(N_c)$ Seiberg duality from the previous section is described in Fig \ref{seibergfig}. 

In this work we deal with supersymmetric (with four supercharges) dualities of quiver gauge theories built from bifundamental matter, namely the matter content of gauge theories are represented as bifundamentals between gauge groups.

\subsection{Partition functions and corresponding solutions}

In this section we collect the expressions for the matrix models associated to the lens index, supersymmetric index and squashed sphere partition functions in four, three and two dimensions. The localization technique enables us to calculate the partition function of supersymmetric gauge theories with four supercharges on different manifolds exactly
\begin{equation} \label{matrixint}
Z =  \frac{1}{|W|}   \SumInt \frac{dz_i}{2 \pi i z_i} \prod_{i=1}^{\text{rank}G} Z_{gauge} (z_i; m_i)  \prod_{\Phi} Z_{\Phi}(z_i, t_a,; m_i)  .
\end{equation}
In the examples of the next sections we only discuss theories without the Chern-Simons and Fayet-Iliopoulos terms are therefore the classical terms is absent in our expressions.

\begin{itemize}
\item The integral is performed over the Cartan subgroup of the gauge
group. It is parameterized by the diagonal entries of the real scalar
$z$ in the gauge group. 
\item The factor of inverse $|W|$ represents the order of the Weyl group of the gauge group. 

\end{itemize}

\begin{table}[h]
\begin{center}
\def\arraystretch{1.6}
    \begin{tabular}{ | l | l | l | p{5cm} |}
    \hline
    Manifold & Vector Multiplet & Chiral Multiplet \\ \hline
    $S^3/\mathbb{Z}_r \times S^1$ & $\prod_{\alpha} {\left(\Gamma_e ( \alpha(z)^{\pm}; \pm \rho_j(m); p,q)\right)^{-1}}
$ &  $\prod_{j }\Gamma_e(
(pq)^{\frac{\Delta_j}{2}+\rho_j(m)+\phi_j(n)} \rho_j(z)\phi_j(a); p,q)$ \\ \hline
$S^3\times S^1$ & $\prod_{\alpha} \left(\Gamma\left(\alpha(z)^{\pm};p,q \right)\right)^{-1}
$ &  $\prod_{j} \Gamma(
(pq)^\frac{\Delta_j}{2} \rho_j(z)\phi_j(a);p,q)$  \\ \hline
$S^2\times S^1$ & $ \prod_{\alpha} q^{-\frac{1}{2}|\alpha(m)|} (1-{\alpha(z)}^{\pm}q^{\frac{|\alpha(m)|}{2}}) $  &  $ \prod_j \frac{(  q^{1-\frac{\Delta_j}{2}+\frac{|\rho_j(m)+\phi_j(n)|}{2}} \rho_j(z)^{-1}\phi_j(a)^{-1};q) _\infty}{(q^{\frac{\Delta_j}{2}+\frac{|\rho_j(m)+\phi_j(n)|}{2}} \rho_j(z)\phi_j(a);q)_\infty}$  \\ \hline
$S_b^3/\mathbb{Z}_r$ & $ \prod_{\alpha} \left(\hat s_{b, \alpha(m)} 
\left( i\frac{Q}{2}\pm \alpha(z)\right) \right)^{-1}$ &  $\prod_{j}
\hat s_{b,-\rho_j( m)-\phi_j( n)} {\footnotesize 
( i \frac{Q}{2} (1-\Delta_j)-\rho_j(z)-\phi_j(a)) }$  \\ \hline
    $S_b^3$ & $\prod_{\alpha}\left(\gamma^{(2)}\left(\alpha(z)^{\pm}; \omega_1, \omega_2\right) \right)^{-1}$ &  $\prod_{j} \gamma^{(2)}\left(
\frac{(\omega_1+\omega_2)}{2} \Delta_j + \rho_j(z)+\phi_j(a); \omega_1, \omega_2\right)$ \\ \hline
$S^2$ & $e^{2\pi i \delta(m)}\prod_{\alpha}\Big(\frac{\alpha(m)^2}{4}+\alpha(z)^2\Big)
$ &  ${\prod_{j}\frac{\Gamma(\frac{\Delta_j}{2}-i\rho_j(z)-i \phi_j(a)-\frac{\rho_j(m)+\phi_j(n)}{2})}{\Gamma(1-\frac{\Delta_j}{2}+i\rho_j(z)+i\phi_j(a)+\frac{\rho_j(m)+\phi_j(n)}{2})}  }$  \\ \hline
$S^1\times S^1$ & $\prod_{\alpha} \left(\Delta\left(\alpha^{\pm}(z); q,t \right) \right)^{-1}$ &  $\prod_{j} \Delta(
t^\frac{\Delta_j}{2} \rho_j(z)\phi_j(a); q,t)$ \\ \hline 
\end{tabular}    
\end{center}
\caption{Contributions of  vector and chiral  supermultiplets to the supersymmetric partition functions}
\end{table}

\begin{itemize}

\item The contribution of the vector multiplet is parameterized by the positive roots of the algebra. Actually, the Vandermonde determinant in the measure exactly cancels the one loop determinant of the vector multiplet.  
\item The contribution of the matter multiplet  corresponds to the contribution of the $j$-th chiral multiplet with $R$ charge $\Delta_j$. Each chiral multiplet is in the corresponding representation of the gauge group $G$ with weight $\rho_j(z)$ and in the corresponding representation of the flavor group $F$, with weight $\phi_j(a)$. 

\end{itemize}

The observation of the correspondence between supersymmetric theories and solvable lattice models is based on the fact that in both fields appears same special functions of hypergeometric type. For instance, the partition functions of supersymmetric theories on different manifolds can be expressed in terms of the following hypergeometric functions (for details see, e.g. \cite{Gahramanov:2015tta,Gahramanov:gka,Yamazaki:2013fva})

\begin{itemize}

\item $S^3 \times S^1$, $S^3/\mathbb{Z}_r \times S^1$ : elliptic hypergeometric integral 

\item $S^2 \times S^1$ : basic hypergeometric integral

\item $S_b^3$, $S_b^3/\mathbb{Z}_r$ : hyperbolic hypergeometric integral

\item $S^2$, $S^1 \times S^1$: ordinary hypergeometric integral 

\end{itemize}


\section{Integrability from duality}

In this section, we summarize the present status of the gauge/YBE correspondence and list solutions to the star-triangle relation found (or related) via this correspondence. 

A central phenomenon in the construction of integrable lattice models via the gauge/YBE correspondence is the existence of the corresponding Seiberg duality. Here we consider the special Seiberg duality for supersymmetric theories with four supercharges in different dimensions. All known solutions to the star-triangle relation found via the gauge/YBE correspondence results from the following duality \cite{Seiberg:1994pq}

\begin{itemize}

\item \textbf{Theory A}: $SU(2)$ gauge group with $N_f=6$ flavors, chiral
multiplets in the fundamental representation of the flavor group $SU(6)$ and in the
fundamental representation of the gauge group.\\[-0.2cm]

\item \textbf{Theory B}: without gauge degrees of freedom and the chiral fields
(gauge-invariant ``mesons'') in the 15-dimensional totally antisymmetric tensor
representation of the flavor group.

\end{itemize}

Note that we consider only the field content of the dual theories. Of course, this duality has different features in different dimensions, but such details are not crucial for our discussions. For instance, in order to get the right duality, one needs to specify the exact form of the superpotential \cite{Aharony:2013dha,Aharony:2017adm}.

It turns out that identity of partition functions of dual theories can be written\footnote{One needs to break  the flavor symmetry down to $SU(2) \times SU(2) \times SU(2)$.} in the form of the star-triangle relation

\begin{align}\label{Str}\nonumber
\SumInt_{\sigma_0}\mathcal{S}( \sigma _0)\mathcal{W}_{ \eta - \alpha}( \sigma _i , \sigma _0)&\mathcal{W}_{ \eta - \beta}( \sigma _j, \sigma _0) \mathcal{W}_{ \eta - \gamma}( \sigma _k, \sigma _0) \\
 = & \mathcal{R}(\alpha,\beta,\gamma)\mathcal{W}_{  \alpha}( \sigma _j, \sigma _k)W_{ \beta}( \sigma _i, \sigma _k) \mathcal{W}_{  \gamma}( \sigma _j, \sigma _i) \;.
\end{align}
In all models we discuss here the scalar factor $R$ can be absorbed into the Boltzmann weights. In Fig \ref{fig:super} we described the quiver diagram of the above duality which gives the star-triangle relation. In Fig \ref{fig:super} $SU(2)$ gauge group lives on each node presented by circles and bifundamental matter on each edge, the boxes represents the flavor groups $SU(2)$.

\begin{figure}[h] 
    \centering
    \includegraphics[width=0.7\textwidth]{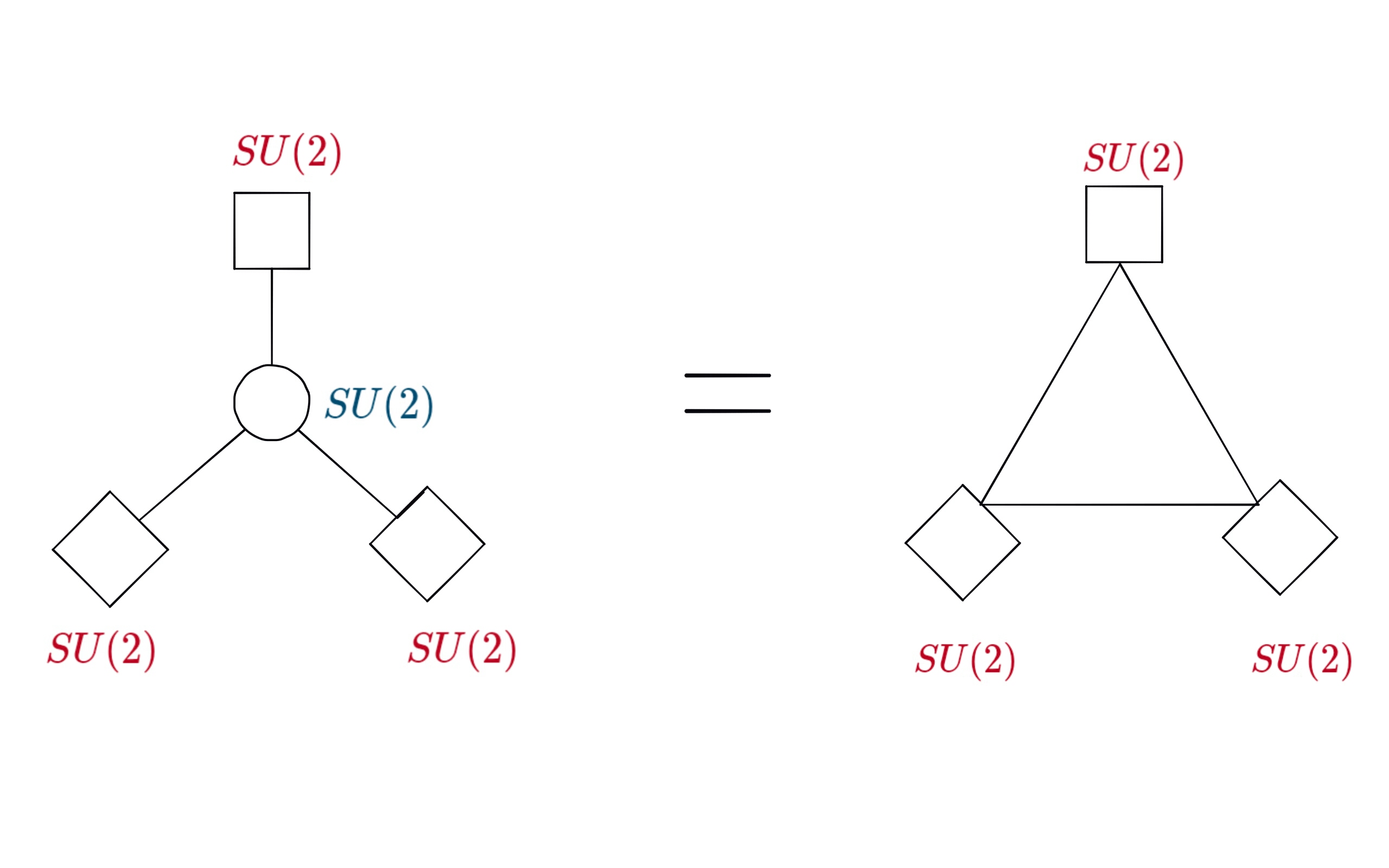}\
   \caption{Supersymmetric duality}
   \label{fig:super}
\end{figure}

It is not clear in gauge/YBE correspondence why the above duality is special, but it is the duality for which the correspondence takes place. We think that one can find solutions to the star-triangle relation only via this duality, all other supersymmetric dualities may give a solution to the star-star relation.



In the context of gauge/YBE correspondence the spin lattice
models can be identified with the quiver gauge theory with SU(2) gauge groups on the sites of the lattice. Then the partition function of the corresponding integrable model is equivalent to the supersymmetric partition function of the corresponding supersymmetric quiver gauge
theory. The contribution of chiral and vector multiplets to the supersymmetric partition function correspond to the nearest-neighbor Boltzmann weights and the self-interaction, respectively. In Fig \ref{fig:super} we describe the correspondence of partition functions pictorially.

\begin{figure}[h]
    \centering
    \includegraphics[width=1.00\textwidth]{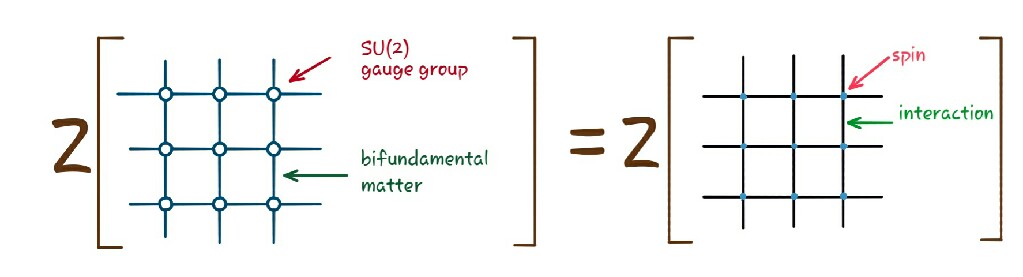}
     \caption{Equivalence of the partition functions in the context of gauge/YBE correspondence: The left-hand side is the partition function of the supersymmetric quiver gauge theory, and the right side is the partition function of the integrable lattice model.}
   \label{fig:super}
\end{figure}

We would like to mention that the inversion relation (\ref{In-rel2}) has an interesting counterpart on supersymmetry side of the gauge/YBE correspondence, namely, it is related to the chiral symmetry breaking of the corresponding supersymmetric gauge theory. Such relation can be derived in many different ways, from supersymmetric gauge theory side one can obtain inversion relation by an accurate limit of parameters in the partition functions of dual theories \cite{Spiridonov:2014cxa}.

The Seiberg duality can be realized in the context of the brane language \cite{Elitzur:1997fh,Elitzur:1997hc,Giveon:1998sr}, namely one can obtain the duality by exchanging NS5-branes. Such construction gives an opportunity to obtain integrable lattice models directly using brane construction. We will not discuss this direction in the paper and refer the interested reader to the papers \cite{Yamazaki:2012cp,Yagi:2015lha,Yagi:2016oum}.

We finish this section by remarking that one can also construct an IRF-type integrable multispin model. For that one needs to take $SU(N)$ gauge theory with $SU(N) \times SU(N)$ flavor symmetry. Then the identity of partition functions for dual theories can be written as the star-star relation for the IRF-type model. Unfortunately, it is quite complicated to prove analytically such integral identities \cite{Kels:2017toi,Bazhanov:2013bh,Bazhanov:2011mz,Gahramanov:2017idz,Yamazaki:2015voa}.

Below we list all solutions to the star-triangle relation found (or related) by gauge/YBE correspondence, but we do not treat them in detail. 

\subsection{$S^3/\mathbb{Z}_r \times S^1$ partition function and solution}

The solution to the star-triangle relation was found\footnote{Kels also gives an analytic proof of the integral identity in \cite{Kels:2015bda}.} by Kels in \cite{Kels:2015bda} on the bases of the special case of the star-star relation found by Yamazaki\footnote{Actually, Yamazaki constructed \cite{Yamazaki:2015voa} the star-star relation for the dual theories with $SU(N)$ gauge group and $SU(N) \times SU(N)$ flavor group.} in \cite{Yamazaki:2013nra}. 

The idea is that the sum-integral identity for the  four-dimensional lens supersymmetric indices of supersymmetric dual theories can be written as the star-triangle relation

\begin{align}\nonumber
\frac{(p^{r},p^{r})_{\infty}(q^{r},q^{r})_{\infty}}{2}\sum_{m_0=0}^{r-1} \int_{0}^{1}dz&\frac{\prod_{i=1}^{6}\Gamma_e\Big((a_i\pm z),(m_i\pm m_0);\sigma,\tau\Big)}{\Gamma_e(\pm 2z,\pm 2m_0);\sigma,\tau}\\ 
&\qquad\qquad=\prod_{1\leq i<j\leq 6} \Gamma_e\Big((a_i+a_j),(m_i+m_j);\sigma,\tau\Big) ,
\end{align}
 with the  balancing conditions $\sum_{i=1}^6 a_i=\sigma+\tau$ and $\sum_{i=1}^6 m_i=0$. The definition of the lens gamma function is given in Appendix \ref{App-elg}.
 

The lattice model has two spin variables on each site, discrete and continuous spin 
\begin{equation}
\sigma_j=(x_j, m_j), \;\; \text{where} \;\; 0 \leq x_j < 2\pi, \;\; m_j=0,1,2,\dots ,[r/2] \;.
\end{equation}
The Boltzmann weight and self-interaction term of the model are \cite{Kels:2015bda}. 
\begin{align}\nonumber
 \mathcal{W}_{ \alpha } ( \sigma _i, \sigma _j)  = \frac{e^{-2 \alpha ([m_i-m_j]_{ \pm }+[m_i+m_j]_{\pm})/r}}{k( \alpha )} & \frac{{\Phi _{r, m_i-m_j}(x_i-x_j+i \alpha )}}{{\Phi _{r, m_i-m_j}(x_i-x_j-i \alpha )}}\\
 &\quad\times\frac{\Phi _{r,m_i+m_j}(x_i+x_j+i \alpha )}{\Phi _{r,m_i+m_j}(x_i+x_j-i \alpha )},
 \end{align}
\begin{align}
     \mathcal{S}(\sigma_0) & = \frac{ \varepsilon_0}{\pi}(p^{2r};p^{2r})_{\infty}(q^{2r};q^{2r})_{\infty}e^{2\eta[2m_0]_{\pm}/r} \Phi_{r,-2m_0}(-2x_0-i\eta) \Phi_{r,2m_0}(2x_0-i\eta),
  \end{align}
where $k( \alpha )$ represents the partition function per edge and is defined as follows:
 \begin{equation}
k(\alpha)=\exp \Big( \sum_{n \neq 0} \frac{e^{4\alpha n}((pq)^{rn}-(pq)^{-rn})}{n((pq)^{2n}-(pq)^{-2n})(p^{rn}-p^{-rn})(q^{rn}-q^{-rn})}  \Big),
\end{equation}  
and   
  \begin{align}
  \varepsilon_0= \begin{cases}
  \frac{1}{2}, & \text{if } m_0=0 \,\, \text{or } [r-m_0]_r\\
            1, & \text{otherwise } .
    \end{cases}
  \end{align}
Here we use the notation of \cite{Kels:2015bda} for the lens elliptic gamma function which is defined in the following way 
  \begin{align}
  \Phi _{r,m}(z)=\prod_{j,k=0}^{\infty}\frac{1-e^{2iz}p^{-2[m]_r}(pq)^{2j+1}p^{2kr+2r}}{1-e^{-2iz}p^{2[m]_r}(pq)^{2j+1}p^{2kr}} \frac{1-e^{2iz}q^{-2[m]_r}(pq)^{2j+1}q^{2kr}}{1-e^{-2iz}q^{-2[m]_r}(pq)^{2j+1}q^{2kr+2r}},
  \end{align}
where $[m]_r \in \{0,1,2\dots, r-1\}$ and $[m]_{\pm}:= [m]_{+}[m]_{-}$.

We should mention that the solution in terms of lens gamma functions is the top level known solution to the star-triangle relation at this time which gives almost all known other models in the limiting case.


\subsection{$S^3 \times S^1$ partition function and solution}

The identity for the superconformal indices of the dual theories is the following elliptic beta integral \cite{Dolan:2008qi}
\begin{align} 
\frac{(q;q)_\infty(p;p)_\infty}{2} \oint \frac{dz}{2\pi i z}  \,\frac{\prod_{i=1}^6\Gamma(a_iz^{\pm};p,q)}{\Gamma(z^{\pm 2};p,q)} =\!\!\!\!\prod_{1\leq i<j\leq6}\Gamma(a_ia_j;p,q)\,,
\end{align}
with the balancing condition $\prod_{i=1}^{6} a_i = p q$. This identity was introduced and proven by Spiridonov in \cite{Spiridonovbeta}. Later Bazhanov and Sergeev interpreted this identity as the star-triangle relation and introduced a new integrable spin lattice model \cite{Bazhanov:2010kz}. 

In the corresponding integrable lattice model spin variables get continuous values
\begin{equation}
 0 \leq \sigma_i< 2\pi,
 \end{equation}
and the Boltzmann weights are expressed in terms of elliptic gamma functions
\begin{align}
 \mathcal{W}_{\alpha}(\sigma_i,\sigma_j) & = \frac{1}{k(\alpha)}\Gamma(e^{\alpha-\eta\pm i\sigma_i\pm i\sigma_j};p,q),
 \end{align}
\begin{equation}
 \mathcal{S}(\sigma_0)  =  \frac{(p;p)_ \infty(q;q)_ \infty  }{4 \pi }  \theta(e^{\pm 2i\sigma_0};q),
\end{equation}
where
\begin{align}
k(\alpha)=\frac{\Gamma(e^{2\alpha}(pq)^2;p,q,(pq)^2)}{\Gamma(e^{2\alpha}pq;p,q,(pq)^2)}\,\,
\; \; \text{and} \;\; \Gamma(z;p,q,t):=\prod_{i,j,k=0}^{\infty}\frac{1-z^{-1}p^{i+1}q^{j+1}t^{k+1}}{1-zp^iq^jt^k}.
\end{align}

Here we use the notations of \cite{Spiridonov:2010em}. In order to keep track of notations of the paper \cite{Bazhanov:2010kz} by Bazhanov one has to use  the following form of the elliptic gamma  function:
 \begin{align}\nonumber
  \Phi (x) =  \Gamma (e^{-i(x-i \eta )};p^2,q^2);  
\end{align} 
where $p=e^{i\pi\tau}$, $q=e^{i\pi\sigma}$, $\eta=-i\pi(\tau+\sigma)$, $\operatorname{Im}\tau>0$ and $\operatorname{Im}\sigma>0$.Then the Boltzmann weights for the Bazhanov-Sergeev model get the following form:
\begin{equation}
      \mathcal{W}_ \alpha (\sigma_i,\sigma_j)=k( \alpha )^{-1} \frac{ \Phi (\sigma_i-\sigma_j+i\alpha)}{ \Phi (\sigma_i-\sigma_j-i\alpha)} \frac {\Phi (\sigma_i+\sigma_j+i \alpha )}{ \Phi (\sigma_i+\sigma_j-i \alpha )}, 
\end{equation} 
  \begin{equation}
       S(\sigma_0)= \frac{e^ {\eta /4}}{4 \pi } \vartheta _1(\sigma_0| \tau ) \vartheta _1(\sigma_0|  \sigma  ),
  \end{equation}
where 
\begin{equation}
k(\alpha)= \exp \Big ( \sum_{n \neq 0} \frac{e^{2\alpha n}}{n(p^n-p^{-n})(q^n-q^{-n})((pq)^n+(pq)^{-n})} \Big) .
\end{equation}
Here  $\vartheta _1(x| \tau )$ is the Jacobi theta function defined in the Appendix \ref{App-Theta}.

This solution of the star-triangle relation can be obtained from the lens supersymmetric index by taking $r=1$. Note that in \cite{Bazhanov:2013bh} the authors found a multi-spin generalization\footnote{ In terms of supersymmetric gauge theories one needs to consider the dual theories with $SU(N)$ gauge group and $SU(N) \times SU(N)$ flavor group.} of this model and constructed the star-star relation for it.

\subsection{$S^2 \times S^1$  partition function and solution}


The identity for $S^2 \times S^1$  partition functions (three-dimensional supersymmetric indices) of daul theories\footnote{Note that here we presented the so-called generalized supersymmetric index \cite{Kapustin:2009kz}. The ordinary supersymmetric index with enhanced symmetry for the $N_f=4$ case was considered in \cite{Gahramanov:2013xsa} (see also the cases with broken gauge group in \cite{Gahramanov:2013rda,Gahramanov:2014ona}).} is the following $q$-beta hypergeometric sum-integral \cite{Gahramanov:2016wxi}
\begin{align}\nonumber
     \sum_{m=-\infty}^{\infty}  \oint & \prod_{i=1}^6 \frac{(q^{1+(m+n_i)/2}/a_iz;q)_\infty(q^{1+(n_i-m)/2} z/a_i ;q)_\infty}{(q^{(m+n_i)/2}a_iz;q)_\infty(q^{(n_i-m)/2} a_i/z ;q)_\infty} \frac{(1-q^m z^{\pm 2})}{q^m z^{6m}} \frac{dz}{2\pi i z} \\  &\qquad\qquad\qquad\qquad\qquad=\frac {2}{ \prod_{i=1}^6 a_i^{n_i}}  \prod_{1 \leq i<j  \leq 6} \frac {(q^{1+(n_i+n_j)/2}/a_i a_j;q)_\infty}{(q^{(n_i+n_j)/2}a_i a_j;q)_\infty},
\end{align}
with the balancing conditions $\prod_{i=1}^6 a_i = q$ and $\sum_{i=1}^6 n_i =  0$. This identity was studied in the context of supersymmetric dualities in \cite{Gahramanov:2016wxi}, integrability in \cite{Gahramanov:2015cva,Kels:2015bda}, and from point of view of orthogonal polynomials in \cite{Rosengren:2016mnw}.

In the corresponding integrable model we again have discrete and continuous spin variables
\begin{equation}
\sigma_j=(x_j,m_j) \;\; \text{where} \;\;  0 \leq x_j < 1 \; \text{and} \;    m_j \in \mathbb{Z}.
\end{equation}
The Boltzmann weights for the model are
\begin{align} \nonumber
        \mathcal{W}_{\alpha}(\sigma_i,\sigma_j) & =\frac{q^{-2i(x_im_i+x_jm_j)}} {k(\alpha)}\frac{(q^{1+(m_i+m_j)/2}q^{\eta-\alpha-i(x_i+x_j)};q)_\infty}{(q^{(m_i+m_j)/2}q^{\alpha-\eta+i(x_i+x_j)};q)_\infty}   \frac{(q^{1+(m_j-m_i)/2}q^{\eta-\alpha+i(x_i-x_j)};q)_\infty}{(q^{(m_j-m_i)/2}q^{\alpha-\eta-i(x_i-x_j)};q)_\infty} \\ \label{W}
 & \quad \times   \frac{(q^{1-(m_i+m_j)/2}q^{\eta-\alpha+i(x_i+x_j)};q)_\infty}{(q^{-(m_i+m_j)/2}q^{\alpha-\eta-i(x_i+x_j)};q)_\infty}    \frac{(q^{1+(m_i-m_j)/2}q^{\eta-\alpha+i(x_j-x_i)};q)_\infty}{(q^{(m_i-m_j)/2}q^{\alpha-\eta-i(x_j-x_i)};q)_\infty}, 
\end{align}

\begin{equation}
        \mathcal{S}(\sigma_0)= \frac{1}{2\pi q^m}\frac{(q^{\pm 2x_0+m};q)_\infty}{(q^{\pm 2x_0+m+1};q)_\infty},
\end{equation}
where
\begin{align}\label{qp-Poch}
k(\alpha)=\frac{(q^2e^{4\pi i \alpha},qe^{-4\pi i \alpha};q,q^2)_\infty}{(qe^{4\pi i \alpha},q^2e^{-4\pi i \alpha};q,q^2)_\infty};\,\,\,\,\,\,
 (z,w;p,q)_{\infty}:=\prod_{i,j=0}^{\infty}(1-zp^iq^j)(1-wp^iq^j).
\end{align}
   
The multi-spin generalization of this solution  was constructed by authors in \cite{Gahramanov:2017idz}.

In \cite{Kels:2015bda} Kels uses slightly different notations. In his case the  Boltzmann weight and self interaction terms are expressed as
\begin{align}\nonumber
&\mathcal{W}_{\alpha}(\sigma_i.\sigma_j)=\frac{e^{-2\alpha|m_i-m_j|-2\alpha|m_i+m_j|}}{k(\alpha)}\frac{Q(x_i-x_j+i\alpha,m_i-m_j)}{Q(x_i-x_j-i\alpha,m_i-m_j)}\\
&\qquad\qquad\qquad\qquad\qquad\qquad\qquad\qquad\qquad\qquad\times
\frac{Q(x_i+x_j+i\alpha,m_i+m_j)}{Q(x_i+x_j-i\alpha,m_i+m_j)},
\end{align}
\begin{align}
\mathcal{S}(\sigma_0)=\frac{1}{2\pi}Q(2x_0-i\eta,2m_0)Q(-2x_0-i\eta,-2m_0),
 \end{align}
with
\begin{align}
k(\alpha)=\exp\Big\{-\sum_{n\neq 0} \frac{e^{4\alpha n}}{n((\boldsymbol{q})^{n}-(\boldsymbol{q})^{-n})}\Big\};
\,\,\,\,\,  Q(z;n)=\frac{(e^{2iz} (p/q)^{-n} (pq)^{1+|n|};(pq)^2)_\infty}{(e^{-2iz} (p/q)^n (pq)^{1+|n|};(pq)^2)_\infty}. 
\end{align}


\subsection{$S_b^3/\mathbb{Z}_r$ partition function and solution}

The sum-integral identity for the  three dimensional  lens partition function of supersymmetric dual theories reads as
\begin{align}\nonumber
\sum_{m_0=0}^{r-1}\int _{\mathbb{R}} \frac{dx_0}{r\sqrt{\omega_1\omega_2}}2\sinh\frac{2\pi}{r\omega_1}(x_0-i\omega_1m_0)\sinh\frac{2\pi}{r\omega_2}(x_0+i\omega_2m_0)\qquad \,\,\,\, \,\,\,\, \,\,\,\,\,\,\,\, \,\,\,\, \,\,\,\, \,\,\,\, \,\,\,\, \,\,\,\,\\
 \times \prod_{i=1}^{6}\frac{\hat s_{b,-m_0-m_k}(x_0+x_k+iQ/2)}{\hat s_{b,-m_0+m_k}(x_0-x_k-iQ/2)}=\prod_{1\leq i<j\leq 6} \hat s_{b,-m_j-m_k}(x_j+x_k+iQ/2), 
\end{align}
with the balancing conditions $i\sum_{i=1}^{6}x_i=Q$ and $\sum_{i=1}^{6}m_i=0$, where $Q=b+\frac{1}{b}$.
Here we used the improved double sine function defined as
\begin{align}
\hat s_{b,-m}(x)=e^{\frac{i\pi}{2r}([m](r-[m])-(r-1)m^2)} \varphi_{r,m}(x) \;.
\end{align}
The function $\varphi_{r,m}(z)$ is generalization of the Faddeev's quantum dilogarithm
\begin{align}\nonumber
 \varphi_{r,m}(z) =\exp\Big\{\int_0^{\infty} dx \Big(\frac{iz}{\omega_1  \omega _2 r x^2}- &\frac{\sinh(2izx-\omega_1(r-2[m]x))}{2x\sinh(\omega_1 rx)\sinh((\omega_1 + \omega _2) x) } \\ 
 &\quad-\frac{\sinh(2izx+\omega_2(r-2[m]x))}{2x\sinh(\omega_2 rx)\sinh((\omega_1 + \omega _2) x) }   \Big) \Big\}.
\end{align}
The hyperbolic limit of the lens index solution was presented in \cite{Gahramanov:2016ilb} where the authors  also give an interpretation of this solution in terms of supersymmetric gauge theory.

In this model there are continuous and discrete spin variables living on each site of the lattice 
\begin{equation}
\sigma_j=(x_j,m_j) \;\; \text{where} \;\; 0 \leq x_j < \infty \; \text{and} \;
    m_j=0,1,2,\dots,\lfloor r/2 \rfloor \;,
\end{equation}
and the Boltzmann weights are
\begin{align}
   \mathcal{W}_{ \alpha } ( \sigma _i, \sigma _j) & = \frac{1}{k(\alpha)} \frac{{\varphi _{m_i+m_j}(x_i+x_j+i \alpha )\varphi _{m_i-m_j}(x_i-x_j+i \alpha )}}{{\varphi _{m_i+m_j}(x_i+x_j-i \alpha )\varphi _{m_i-m_j}(x_i-x_j-i \alpha )}},\\
\mathcal{S}( \sigma _0) & = \frac{4\varepsilon_0}{ r\sqrt{ \omega _{1} \omega _{2}}}  \sinh\Big( \frac{ 2\pi }{ \omega _1r}(x_0-i \omega _1m_0) \Big)\sinh\Big(\frac{ 2\pi }{ \omega _2r}(x_0+i \omega _2m_0)\Big),
\end{align}
where the normalization constant gets the following form
\begin{align}
   k (\alpha) & = \exp\Big\{\int_0^{\infty} dx \Big(-\frac{\alpha}{\omega_1  \omega _2 r x^2}+\frac{\sinh(4\alpha x)\sinh(2r\eta x)}{2x\sinh(\omega_1 rx)\sinh(\omega_2 rx)\sinh(4\eta x)} \Big)       \Big\}.
\end{align}

For $r=1$ the this solution reduces to the Spiridonov's generalization \cite{Spiridonov:2010em} of the Faddeev-Volkov model which we will consider in the next section.


\subsection{$S_b^3$ partition function and solution}


The hyperbolic hypergeometric integral identity for the  three-dimensional  squashed sphere partition function\footnote{Note that depending on different squashings of three-sphere and on choice of the preserved charges in the supersymmetric localization one can get different partition functions depending on values of squashing parameter $b$ (see \cite{Imamura:2013nra,Hosomichi:2014hja,Nian:2013qwa} for details). For the specific choice of the preserved supercharge the partition function on squashed three-sphere with $SU_l(2) \times U_r(1)$ isometry gives the $b=1$ case \cite{Hama:2011ea} (the case $b=1$ also corresponds to the round sphere partition function). The integral identity with general values of $b$  is written for the partition functions of dual theories on three-sphere with $U(1) \times U(1)$ isometry (also with $SU_l(2) \times U_r(1)$ \cite{Imamura:2011wg}).} of supersymmetric dual theories reads as
\begin{equation}\label{eq:gt_integral}
  \int_{-\infty}^\infty
  \frac{\prod_{j=1}^6\gamma^{(2)}(g_k\pm i z;\omega) }
  {\gamma^{(2)}(\pm 2 i z;\mathbf{\omega}) } d z
  =2\sqrt{\omega_1\omega_2}
  \prod_{1\leq j<k\leq 6}\gamma^{(2)}(g_j+g_k;\mathbf{\omega}) \;,
\end{equation}
with the balancing condition $\sum_{j=1}^{6} g_{j} = \omega_{1} + \omega_{2} := 2\eta$. The definition of the hyperbolic gamma function is given in Appendix \ref{App-hypg}. 

This integral identity for supersymmetric dual theories was computed in \cite{Teschner:2012em} by applying the reduction procedure of \cite{Dolan:2011rp} to the elliptic beta integral. As a star-triangle relation this identity was considered by Spiridonov in \cite{Spiridonov:2010em}. It is a generalization of the Faddeev-Volkov model\footnote{Actually from the supersymmetric viewpoint the Faddeev-Volkov model corresponds to above duality with broken gauge group.} \cite{Faddeev:1994fw,Bazhanov:2007mh,Bazhanov:2007vg}. In this integrable model spins get continuous values and the Boltzmann weights are
\begin{align} \label{Spirsol1}
\mathcal{W}_{\alpha}(\sigma_i,\sigma_j) & =\frac{1}{k(\alpha)}\gamma^{(2)}(\alpha-\eta\pm i\sigma_i\pm i\sigma_j;\omega),\\ \label{Spirsol2}
 \mathcal{S}(\sigma_0) & =\frac{2}{\sqrt{\omega_1\omega_2}} \sinh\frac{2\pi \sigma_0 }{\omega_1}\sinh\frac{2\pi \sigma_0} {\omega_2},
\end{align}
where
\begin{align} \nonumber
k(\alpha)=\exp\Big\{-\pi i \alpha^2-\frac{\pi i}{24}(1-2(b+b^{-1})^2)\Big\} &\frac{(\tilde{q}\, e^{2\pi i u/b};\tilde{q}^2)_\infty}{(q\,e^{2\pi iu b};q^2)_\infty}\\
\qquad\qquad\qquad\times &\prod_{j,k=0}^{\infty}\frac{1+e^{\pi i u/(b+b^{-1})}\tilde{p}^{j+1}\tilde{q}^{2k}}{1-e^{\pi i u/(b+b^{-1})}\tilde{p}^{j+1}\tilde{q}^{2k}},
\end{align}
with $b=\omega_1$ and $b^{-1}=\omega_2$, $q=e^{2\pi i b^2}$, $\tilde{q}=e^{-2\pi i /b^2}$, and $\tilde{p}=e^{-\pi i/(1+b^2)}$.

This solution can also be  expressed in a different way \cite{Bazhanov:2016ajm}
\begin{align}
\mathcal{W}_{\alpha}(\sigma_i,\sigma_j) & =\frac{1}{k(\alpha)}\frac{\Phi(\sigma_i-\sigma_j+i\alpha)}{\Phi(\sigma_i-\sigma_j-i\alpha)} \frac{\Phi(\sigma_i+\sigma_j+i\alpha)}{\Phi(\sigma_i+\sigma_j-i\alpha)},\\
\mathcal{S}(\sigma_0) & = 2 \sinh(2\pi \eta \sigma_0) \sinh(2\pi \eta^{-1} \sigma_0),
\end{align}
with
\begin{equation}
k(\alpha)=\exp\Big\{\frac{1}{8}\int_{PV}\frac{e^{4\alpha w}}{\sinh ({\eta w})\sinh ({\eta^{-1} w})\cosh((\eta+\eta^{-1})w)}\frac{dw}{w}\Big\},
\end{equation}
where the other version of the hyperbolic gamma function is used
\begin{align}
\Phi(z)=\exp\Big\{\frac{1}{4}\int_{PV}\frac{e^{-2izw}}{\sinh (wb)\sinh (wb^{-1})}\frac{dw}{w}\Big\} \;.
\end{align}

One can find the proof of the star-triangle relation in terms of Boltzmann weights (\ref{Spirsol1})-(\ref{Spirsol2}) in many places, e.g. see \cite{Bultthesis}. The quasi-classical limit $b \rightarrow 0$ of the model was considered in \cite{Bazhanov:2016ajm}.


\subsection{$S^2$ partition function and solution}

The solution to the star-triangle relation was obtained by Kels in \cite{Kels:2013ola} (see also \cite{Kels:2015bda}). The interpretation in terms of supersymmetric sphere partition function and the star-star relation is given in 

The integral identity for sphere partition functions of dual theories reads as
\begin{align}\nonumber
  \sum_{m\in \mathbb{Z}}\int_{-\infty}^{+\infty} \frac{dz}{2\pi i z} \frac{\Gamma(m\pm 2iz+1)}{\Gamma(m\pm2iz)}  \prod_{i=1}^{6} &\frac{\Gamma(\frac{m+n_i}{2}+a_i+iz)}{\Gamma(1+\frac{m+n_i}{2}-a_i-iz)} \frac{\Gamma(\frac{m-n_i}{2}+a_{i}-iz)}{\Gamma(1+\frac{m-n_i}{2}-a_{i}+iz)}\\
\qquad &= \prod_{1\leq i<j\leq6} \frac{\Gamma(a_i+a_j+\frac{n_i+n_j}{2})}{\Gamma(1-a_i-a_j-\frac{n_i+n_j}{2})},
\end{align}
with the balancing conditions $\sum_{i=1}^6 a_i=1$ and $\sum_{i=1}^6 n_i=0$.

In the corresponding models spin variables get discrete and continuous  values 
\begin{equation}
\sigma_j=(x_j,m_j) \;\; \text{where}\quad\,     0 \leq x_j < 2\pi \; \text{and} \;    m_j \in Z \;.
\end{equation}
The Boltzmann weight and self-interaction term for the model are  
 \begin{align}
 \mathcal{W}_{\alpha}(\sigma_i,\sigma_j)=\frac{\Gamma(\frac{1+\alpha}{2})}{\Gamma(\frac{1-\alpha}{2})}\,
\frac{\Gamma(\frac{1-\alpha}{2}\pm\frac{i(x_i+x_j)-(m_i+m_j)}{2})\,\Gamma(\frac{1-\alpha}{2}\pm\frac{i(x_i-x_j)-(m_i-m_j)}{2})}
{\Gamma(\frac{1+\alpha}{2}\pm\frac{i(x_i+x_j)+(m_i+m_j)}{2})\,\Gamma(\frac{1+\alpha}{2}\pm\frac{i(x_i-x_j)+(m_i-m_j)}{2})},
  \end{align}   
\begin{equation}
 \mathcal{S}(\sigma_0)=\frac{1}{2\pi}\frac{\Gamma(m_0\pm 2ix_0+1)}{\Gamma(m_0\pm 2ix_0)}.  
\end{equation}


\subsection{$S^1 \times S^1$ partition function and solution}


Integral identity for the two dimensional supersymmetric indices of dual theory is defined  
\begin{align}\label{s1s1part}
    \frac{1}{2}\Big(\frac{(q;q)^2_{\infty}}{\theta(y;q)}\Big) \oint \frac{dz}{2\pi i z} \frac{\prod_{i=1}^6\Delta(a_iz^{\pm1};q,y)}{\Delta(z^{\pm2};q,y)}
     =\prod_{1\leq i<j\leq 6}\Delta(a_ia_j;q,y),
\end{align}
with the  balancing condition  $ \prod_{i=1}^6 a_i= \frac{q}{y}$. Here we used the following ratios of the theta functions 
\begin{align}
\quad\Delta(z;q,y):=\frac{\theta(zy,q)}{\theta(z,q)}
\end{align}
This integral identity was written as the star-triangle relation in \cite{Jafarzade:2017fsc}. In this model spin variables get the cantionuos values $ 0 \leq \sigma_i< 2\pi$. The Boltzmann weights and self-interaction term are
\begin{equation}
    \mathcal{W}_{\alpha}(\sigma_i,\sigma_j)=\frac{1}{k(\alpha)}\frac{\theta( e^{-\alpha-\eta\mp i(\sigma_i\pm \sigma_j)};q)}{\theta(e^{\alpha-\eta\pm i(\sigma_i\pm \sigma_j)};q)},\qquad\qquad \quad
\end{equation}
\begin{equation}\label{S}
     \mathcal{S}(\sigma_0)=\frac{1}{4\pi}\Big(\frac{(q;q)^2_{\infty}}{\theta(y;q)}\Big)  \frac{\theta(e^{\pm 2i\sigma_{0}};q)}{\theta(e^{-2\eta\pm 2i\sigma_{0}};q)}, \qquad\quad\,\,\,
\end{equation}
where 
\begin{align}
k(\alpha)=\frac{(q^2e^{2\alpha},qe^{-2\alpha};q,q^2)_\infty}{(qe^{2\alpha},q^2e^{-2\alpha};q,q^2)_\infty} \frac{(q^{-1}e^{-2\alpha},e^{2\alpha};q,q^2)_\infty}{(e^{-2\alpha},q^{-1}e^{2\alpha};q,q^2)_\infty}.
\end{align}
and the infinite products in the formula are defined in the second expression of (\ref{qp-Poch}).

Actually one can use the standard Jacobi theta function notions for this solution. In that case the integral identity (\ref{s1s1part}) will be the indices  of dual theories  defined for RR sector\footnote{Because of spectral duality the index defined for RR sector \cite{Benini:2013xpa,Benini:2013nda} is identical with the index defined for the NS-NS sector \cite{Gadde:2013ftv}}. Then the Boltzmann weights have the following form 
\begin{equation}
     \mathcal{W}_{\alpha}(\sigma_i,\sigma_j)=\frac{\theta_1( e^{-\alpha-\eta\mp i(\sigma_i\pm \sigma_j)};q)}{\theta_1(e^{\alpha-\eta\pm i(\sigma_i\pm \sigma_j)};q)},
\end{equation}
\begin{equation}
     \mathcal{S}(\sigma_0)=\frac{1}{4\pi} \Big(\frac{\eta(q)^{3}}{i\theta_1(e^{-2\eta};q)}\Big) \frac{\theta_1(e^{\pm 2i\sigma_0};q)}{\theta_1(e^{-2\eta\pm 2i\sigma_0};q)}.
\end{equation}
We would like to mention that the multi-spin case of the model was constructed in \cite{Yamazaki:2015voa}. 
In \cite{Jafarzade:2017fsc} the authors considered the high temperature limit of this model and obtained a new solvable model with the following Boltzmann weights
\begin{equation}
   \mathcal{W}_\alpha(\sigma_i,\sigma_j)=\frac{\sinh\pi((-\eta+\alpha)\pm i(\sigma_i\pm\sigma_j)+t)}{\sinh\pi((-\eta+\alpha)\pm i(\sigma_i\pm\sigma_j))},\qquad\qquad\qquad\,\,\,\,\,
\end{equation}
\begin{equation}
    S(\sigma_0)=\frac 12 \frac{-\pi}{\sinh\pi t}\frac{\sinh\pi(\pm 2 i\sigma_0) }{\sinh \pi(\pm 2i\sigma_0+t)}, \qquad \qquad\qquad\qquad
    \end{equation}
\begin{equation}
  R(\alpha,\beta,\gamma) =\frac{\sinh\pi(-2\alpha+t)}{\sinh\pi(-2\alpha)}\frac{\sinh\pi(-2\beta+t)}{\sinh\pi(-2\beta)}\frac{\sinh\pi(-2 \gamma+t)}{\sinh\pi(-2\gamma)}\,\,\,\,
\end{equation}
which is the solution of (\ref{Str}) with continuous spin variables.

\section{Summary and discussion}


In this paper, we review integrable Ising-like square lattice models obtained (or related) from supersymmetric gauge theory computations. 

In all models which we consider in the paper the Boltzmann weights depend only on the differences of the spin variables at the neighbor sites and rapidity variables at the ends of the edge. In principle, it should possible to obtain a model without such symmetries from supersymmetric computations \cite{GahJaf2}.

However, many key questions have not been answered. For example, given a supersymmetric duality with a gauge and matter multiplets in some representation of the gauge and flavor groups, what is the corresponding integrable lattice model? In our opinion, this is one of the most important questions posed by consideration of the subject. Actually, it is absolutely unclear whether gauge/YBE is generic or a feature of a few special duality.


The reader might wonder whether the gauge/YBE could be used for  three-dimensional lattice models. Actually, there are a lot of attempts to extend the idea of integrability to three- \cite{Kuniba:2015azm, Bazhanov:2005as,Stroganov:1997br} and higher-dimensional generalization \cite{Bazhanov:1981zm,frenkel1991} of lattice models. The condition of commutativity for the transfer matrices in the three-dimensional case takes the form of the so-called tetrahedron equation by Zamolodchikov \cite{zamolodchikov1980tetrahedra}. It would be interesting to extend the relationship between supersymmetric gauge theory computations and integrable models to higher dimensions and find a solution to the tetrahedron equation. 

We remark that the reader might have expected that one can use integrability methods to study dualities and supersymmetric gauge theories via the correspondence. The closed form expressions for the partition function of the models discussed above may provide insight towards an understanding of supersymmetric quiver gauge theories and dualities for them. This is an important point and much work remains to be done in this direction.

Linear quiver gauge theories can be formulated using brane constructions, hence one can obtain solutions of the Yang-Baxter equation directly from the latter via corresponding topological quantum field theories. We refer to the work by  \cite{Yagi:2015lha} for a discussion on this formulation. 

The Seiberg duality \cite{Seiberg:1994pq} for quiver gauge theories corresponds to the cluster mutation \cite{fomin2002cluster} for cluster algebras. To be more precise, a cluster algebra defined by a quiver and different quivers are related by the so-called cluster mutation (it is also called cluster transformation) and it happens that the mutation action on quivers is exactly the same as Seiberg duality. It would be interesting to find a relation of integrable models to cluster algebras \cite{Yamazaki:2016wnu}.







\textbf{Acknowledgments}\\
The authors would like to thank all participants of ``Quantum integrability'' seminars at the Physics Department of Mimar Sinan Fine Arts University, especially to Deniz Bozkurt, Do\u{g}u D\"{o}nmez and Sinan Sevim for discussions on the subject and on some parts of the paper. The authors are also thankful to Zainab Nazari for fruitful discussions. IG would like to express his sincere thanks to Vyacheslav Spiridonov, Andrew Kels, Hjalmar Rosengren and Masahito Yamazaki for valuable discussions on the subject. Sh.J is thankful to ICTP (Trieste, Italy) postgraduate diploma program for supporting and the Max-Planck Institut f\"{u}r Gravitationsphysik (Albert-Einstein-Institut) for warm hospitality. IG would like to thank The Institut des Hautes Etudes Scientifiques (IHES) and  Bogoliubov Laboratory of Theoretical Physics, JINR, Dubna where some parts of the work were done for the hospitality.

\section*{Apppendix}

\appendix

\section{Notations}
  
For $z\in\mathbb{C}$, $|q|<1,$ we define the infinite $q$-product (also called $q$-Pochhammer symbol)
\begin{equation}
(z;q)_\infty \ := \ \prod_{k=0}^\infty (1-z q^k).
\end{equation}

Denote that we will use the following definition for theta and  gamma functions (Euler, elliptic and hyperbolic) through the paper 
\begin{align}
(a,b;q)_{\infty}:=(a;q)_{\infty}(b;q)_{\infty}\, ;\,\,\,(az^{\pm 1};q)_{\infty}:=(az;q)_{\infty}(az^{-1};q)_{\infty}.
\end{align}

\section{Theta function}\label{App-Theta}

The $\theta(z,q)$ is the theta function defined by
\begin{equation}
\theta(z;p) \ = \ \prod_{i=0}^{\infty} (1-z^{-1}p^{i+1})(1-z p^i) 
\end{equation} 
It is related to the Jacobi theta functions. The first Jacobi theta function which is used in present paper can be expressed in terms of theta function
\begin{align}
\theta_1(\tau|z) = - i q^{1/8} y^{1/2} (q,q)_{\infty} \theta(y^{-1};q)\;\;\;\;  \text{with $y=e^{2 \pi i z}$},\,\, q=e^{2 \pi i \tau}.
\end{align}
As a product form 
\begin{align}
\theta_1(\tau|z) = -i q^{1/8} y^{1/2} \prod_{k=1}^{\infty} (1-q^k)(1-y q^k)(1-y^{-1}q^{k-1}).
\end{align}
Dedekind eta function is
\begin{equation}
\eta(q) = q^{1/24}\prod_{k=1}^{\infty}(1-q^k).
\end{equation}

\section{Elliptic gamma function}\label{App-elg}

The elliptic gamma function is a meromorphic function of three complex variables with double infinite product \cite{Ruijsenaars:1997:FOA} 
\begin{equation}
\Gamma(u;\tau, \sigma) \ = \ \prod_{i,j=0}^{\infty} \frac{1-e^{2 \pi i((1+j) \tau +(1+i)\sigma-u) }}{1-e^{2 \pi i (j\tau+i\sigma+u)}}, \;
\end{equation}
here $u, \sigma, \tau \in \mathbb{C}$ and $\text{Im} \tau, \text{Im} \sigma >0$. It is convenient to do the following reparametrization
\begin{equation}
p=e^{2 \pi i \tau}, \;\;\ q=e^{2 \pi i \sigma}, \;\; z=e^{2 \pi i u},
\end{equation}
and get the following form 
\begin{align}
\Gamma(z;p,q):=\prod_{i,j=0}^{\infty}\frac{1-z^{-1}p^{i+1}q^{j+1}}{1-zp^iq^j},\,\,\,\,\, 
\end{align}
for $|p|,|q|<1$ and $z\in \mathbb{C}^* $.

The elliptic gamma function satisfies many interesting properties such as symmetry under exchange of parameters $p$ and $q$
\begin{equation}
 \Gamma(z;p,q) = \Gamma(z;q,p) \,,
\end{equation}
the functional relations
\begin{align}
 \Gamma(qz;p,q) = \theta(z;p) \Gamma(z;p,q), \\
  \Gamma(pz;p,q) = \theta(z;q) \Gamma(z;p,q) \,, 
\end{align} 
and the reflection property
\begin{equation}
 \Gamma(z;p,q) \; \Gamma(\frac{pq}{z};p,q) = 1  \, . 
 \end{equation}
The elliptic gamma function has zeros at
 \begin{equation}
 z\in(p^{i+1}q^{j+1});\quad (i,j)\in \mathbb{Z}^{\geq0}
 \end{equation}
 poles at
 \begin{equation}
 z\in(p^{-i}q^{-j});\quad (i,j)\in \mathbb{Z}^{\geq0}
 \end{equation}
and the residue
 \begin{equation}
 \text{Res}_{z=1}\Gamma(z;p,q)=-\frac{1}{(p,p)_\infty(q,q)_\infty}.
 \end{equation}
 \vspace{0.3cm}

Th elliptic Gamma function is an automorphic form of degree 1 associated to a 2-cocycle and it has an $SL(3,Z)$ modular property \cite{Felder200044} based on the following relations
\begin{equation}
\Gamma(u+\tau, \tau, \tau+\sigma) \Gamma(u, \tau+\sigma, \sigma) \ = \ \Gamma(u, \tau, \sigma) \; ,
\end{equation}
\begin{align} 
  \Gamma({ \frac{z}{\sigma}};\frac{\tau}{\sigma},\frac{1}{\sigma}) = e^{i \pi Q(z,\tau, \sigma)} \Gamma(\frac{z-\sigma}{\tau}; \frac{1}{\tau},  \frac{\sigma}{\tau}) \; \Gamma( z; \tau,  \sigma) \;   
\end{align}
Note that the elliptic gamma function is related to the Barnes multiple gamma function of order three \cite{friedman2004shintani}. Probably this relationship has connection to its modular property. 

Lens elliptic gamma function is defined as 
\begin{align}\nonumber
\Gamma_{e}  (z,m;\sigma,\tau)=e^{\phi_e(z,m;\sigma,\tau)}&\prod_{i,j=0}^{\infty}\frac{1-z^{-1}p^{-m}(pq)^{i+1}p^{r(j+1)}}{1-zp^{m}(pq)^{i}p^{rj}} \\
&\qquad\qquad\qquad \times\frac{1-z^{-1}q^{-r+m}(pq)^{i+1}q^{r(j+1)}}{1-zq^{r-m}(pq)^{i}p^{rj}},
\end{align}

\begin{align}
\phi_e(z,m;\sigma,\tau)=2\pi i\Big(R_2(z,0;\sigma,\tau)+ R_2(0,m;\frac{1}{2},-\frac{1}{2})-R_2(z,m;\sigma,\tau)  \Big)
\end{align}
\begin{align}
R_2(z,m;\sigma,\tau)=R(z+m\sigma;r\sigma, \sigma+\tau)+R(z+m\sigma;r\sigma, \sigma+\tau)
\end{align}
where $z\in \mathbb{C}$, $m\in\mathbb{Z}$ and $r\in\{0,1,2,\dots\}$.

Here
\begin{align}
R(z;\sigma,\tau)=\frac{B_{3,3}(z,\sigma,\tau,-1)+B_{3,3}(z-1,\sigma,\tau,-1)}{12}
\end{align}
and  the third order Bernoulli polynomial is
\begin{align}\nonumber
B_{3,3}(z,\omega_1,\omega_2,\omega_3)=\frac{z^3}{\omega_1\omega_2\omega_3}-&\frac{3(\omega_1+\omega_2+\omega_3)z^2}{2\omega_1\omega_2\omega_3}+\frac{(\omega_1^2+\omega_2^2+\omega_3^2)+3(\omega_1\omega_2+\omega_1\omega_3+\omega_2\omega_3)z}{2\omega_1\omega_2\omega_3}\\
&\qquad-\frac{(\omega_1+\omega_2+\omega_3)(\omega_1\omega_2+\omega_1\omega_3+\omega_2\omega_3)}{4\omega_1\omega_2\omega_3}
\end{align}
where $z\in \mathbb{C}$ and $\omega_1,\omega_2,\omega_3\in\mathbb{C}	\setminus \{0\}$.


\section{Hyperbolic gamma function and its extensions}\label{App-hypg}

The hyperbolic gamma function is defined as
\begin{equation}
\gamma^{(2)}(u;\omega_1, \omega_2)=e^{-\pi i B_{2,2}(u;\mathbf{\omega})/2} \frac{(e^{2 \pi i u/\omega_1}\tilde{q};\tilde{q})}{(e^{2 \pi i u/\omega_1};q)} \quad \text{with} \quad q=e^{2 \pi i \omega_1/\omega_2}, \quad \tilde{q}=e^{-2 \pi i \omega_2/\omega_1} \; ,
\end{equation}
where  $B_{2,2}(u;\mathbf{\omega})$ is the second order Bernoulli polynomial,
\begin{equation} B_{2,2}(u;\mathbf{\omega}) =
\frac{u^2}{\omega_1\omega_2} - \frac{u}{\omega_1} -
\frac{u}{\omega_2} + \frac{\omega_1}{6\omega_2} +
\frac{\omega_2}{6\omega_1} + \frac 12.
\end{equation}

There is a version of the hyperbolic gamma function, called the double sine function which is defined as follows.
\begin{align}
s_b(z)= e^{-\frac{i\pi z^2}{2}}\frac{\prod_{k=0}^{\infty}(1+e^{2\pi b z}e^{2\pi i b^2(k+\frac{1}{2})})}{\prod_{k=0}^{\infty}(1+e^{2\pi z/b }e^{-2\pi i /b^2(k+\frac{1}{2})})}
\end{align}

The relation between hyperbolic gamma function and double sine functions reads as
\begin{align}
\gamma^{(2)}(u;\omega_1,\omega_2)= e^{-\frac{\pi i }{2}\Big(B_{2,2}(u;\omega_1,\omega_2)-u^2-\frac{1}{4}(\omega_1+\omega_2)^2)+u(\omega_1+\omega_2))\Big)} s^{-1}_b(iu-\frac{i}{2}(\omega_1+\omega_2)),
\end{align}
for $\omega_1=b$, $\omega_2=b^{-1}$ and $u=\frac{1}{2}(b+b^{-1})-iz$.

The reflection identity for a hyperbolic gamma-function is as follows
\begin{equation}
\gamma^{(2)}(z;\omega_1,\omega_2)\gamma^{(2)}(\omega_1+\omega_2-z;\omega_1,\omega_2) = 1,
\end{equation} 
and the asymptotic formulas are
\begin{align}
\lim_{u \rightarrow \infty}
e^{\frac{\pi \textup{i}}{2} B_{2,2}(u;\omega_1,\omega_2)} \gamma^{(2)}(u; \omega_1,\omega_2)
& =  1, \text{ \ \ for } \text{arg }\omega_1 < \text{arg } u < \text{arg }\omega_2 + \pi,  \\
\lim_{u \rightarrow \infty}e^{-\frac{\pi \textup{i}}{2} B_{2,2}(u;\omega_1,\omega_2)} \gamma^{(2)}(u;\omega_1,\omega_2)
& =  1, \text{  \ \ for } \text{arg } \omega_1 - \pi < \text{arg } u < \text{arg }\omega_2.
\end{align}
It has the following useful properties  
\begin{align}
\gamma^{(2)}(z+\omega_2;\omega_1,\omega_2)=2\sin\Big(\frac{\pi z}{\omega_1}\Big)\gamma^{(2)}(z;\omega_1,\omega_2),\\\nonumber
\gamma^{(2)}(z+\omega_1;\omega_1,\omega_2)=2\sin\Big(\frac{\pi z}{\omega_2}\Big)\gamma^{(2)}(z;\omega_1,\omega_2).
\end{align}

The hyperbolic gamma function has zeros at
 \begin{equation}
 z=\omega_1\mathbb{Z}^{\geq1}+\omega_2\mathbb{Z}^{\geq1},
 \end{equation}
 poles at
 \begin{equation}
z=-\omega_1\mathbb{Z}^{\leq 0}-\omega_2\mathbb{Z}^{\leq 0}.
 \end{equation}

There are different notations and modifications of hyperbolic gamma function, relations between some of them can be found in \cite{Spiridonov:2010em}.

\bibliographystyle{utphys}
\bibliography{references}

\providecommand{\href}[2]{#2}\begingroup\raggedright\begin{thebibliography}{100}

\bibitem{McGuire:1964zt}
J.~B. McGuire, ``{Study of Exactly Soluble One-Dimensional N-Body Problems},''
\href{http://dx.doi.org/10.1063/1.1704156}{{\em J. Math. Phys.} {\bfseries 5}
  no.~5, (1964) 622--636}.

\bibitem{Yang:1967bm}
C.-N. Yang, ``{Some exact results for the many body problems in one dimension
  with repulsive delta function interaction},''
\href{http://dx.doi.org/10.1103/PhysRevLett.19.1312}{{\em Phys. Rev. Lett.}
  {\bfseries 19} (1967) 1312--1314}.

\bibitem{Baxter:1972hz}
R.~J. Baxter, ``{Partition function of the eight vertex lattice model},''
  \href{http://dx.doi.org/10.1016/0003-4916(72)90335-1}{{\em Annals Phys.}
  {\bfseries 70} (1972) 193--228}.
[Annals Phys.281,187(2000)].

\bibitem{Baxter:1982zz}
R.~J. Baxter, {\em Exactly {S}olved {M}odels in {S}tatistical {M}echanics}.
\newblock Academic, London,
1982.
\newblock

\bibitem{Jimbo:1989qm}
M.~Jimbo, ``{Introduction to the {Yang-Baxter} Equation},''
\href{http://dx.doi.org/10.1142/S0217751X89001503}{{\em Int. J. Mod. Phys.}
  {\bfseries A4} (1989) 3759--3777}.

\bibitem{Kulish:1980ii}
P.~P. Kulish and E.~K. Sklyanin, ``{On the solution of the Yang-Baxter
  equation},'' \href{http://dx.doi.org/10.1007/BF01091463}{{\em J. Sov. Math.}
  {\bfseries 19} (1982) 1596--1620}.
[Zap. Nauchn. Semin.95,129(1980)].

\bibitem{Ising:1925em}
E.~Ising, ``{Contribution to the Theory of Ferromagnetism},''
\href{http://dx.doi.org/10.1007/BF02980577}{{\em Z. Phys.} {\bfseries 31}
  (1925) 253--258}.

\bibitem{Onsager:1943jn}
L.~Onsager, ``{Crystal statistics. 1. A Two-dimensional model with an order
  disorder transition},''
\href{http://dx.doi.org/10.1103/PhysRev.65.117}{{\em Phys. Rev.} {\bfseries 65}
  (1944) 117--149}.

\bibitem{Baxter:1997tn}
R.~J. Baxter, ``{Star-triangle and star-star relations in statistical
  mechanics},''
\href{http://dx.doi.org/10.1142/S0217979297000058}{{\em Int. J. Mod. Phys.}
  {\bfseries B11} (1997) 27--37}.

\bibitem{au1989onsager}
H.~Au-Yang, J.~H. Perk, {\em et~al.}, ``Onsager’s star-triangle equation:
  master key to integrability,'' {\em Integrable systems in quantum field
  theory and statistical mechanics} (1989) 57--94.

\bibitem{Kulish:1981gi}
P.~P. Kulish, N.~{\relax Yu}. Reshetikhin, and E.~K. Sklyanin, ``{Yang-Baxter
  Equation and Representation Theory. 1.},''
\href{http://dx.doi.org/10.1007/BF02285311}{{\em Lett. Math. Phys.} {\bfseries
  5} (1981) 393--403}.

\bibitem{Fateev:1982wi}
V.~A. Fateev and A.~B. Zamolodchikov, ``{Selfdual solutions of the star
  triangle relations in Z(N) models},''
\href{http://dx.doi.org/10.1016/0375-9601(82)90736-8}{{\em Phys. Lett.}
  {\bfseries A92} (1982) 37--39}.

\bibitem{Kashiwara:1986tu}
M.~Kashiwara and T.~Miwa, ``{A Class of Elliptic Solutions to the Star Triangle
  Relation},''
\href{http://dx.doi.org/10.1016/0550-3213(86)90591-2}{{\em Nucl. Phys.}
  {\bfseries B275} (1986) 121}.

\bibitem{Hasegawa:1990du}
K.~Hasegawa and Y.~Yamada, ``{Algebraic derivation of the broken Z(N) symmetric
  model},''
{\em Physics Letters A} {\bfseries 146} (1990) 387--396.

\bibitem{Gaudin:1990gf}
M.~Gaudin, ``{La relation etoile-triangle d’un modele elliptique Z(N)},''
{\em Journal de Physique I} {\bfseries 1} (1991) 351–361.

\bibitem{vonGehlen:1984bi}
G.~von Gehlen and V.~Rittenberg, ``{$Z(n$) Symmetric Quantum Chains With an
  Infinite Set of Conserved Charges and $Z(n$) Zero Modes},''
\href{http://dx.doi.org/10.1016/0550-3213(85)90350-5}{{\em Nucl. Phys.}
  {\bfseries B257} (1985) 351}.

\bibitem{AuYang:1987zc}
H.~Au-Yang, B.~M. McCoy, J.~H.~H. Perk, S.~Tang, and M.-L. Yan, ``{Commuting
  transfer matrices in the chiral Potts models: Solutions of Star triangle
  equations with genus $> 1$},''
\href{http://dx.doi.org/10.1016/0375-9601(87)90065-X}{{\em Phys. Lett.}
  {\bfseries A123} (1987) 219--223}.

\bibitem{Baxter:1987eq}
R.~J. Baxter, J.~H.~H. Perk, and H.~Au-Yang, ``{New solutions of the star
  triangle relations for the chiral Potts model},''
\href{http://dx.doi.org/10.1016/0375-9601(88)90896-1}{{\em Phys. Lett.}
  {\bfseries A128} (1988) 138--142}.

\bibitem{Faddeev:1994fw}
L.~D. Faddeev, ``{Current - like variables in massive and massless integrable
  models},'' {\em {Quantum groups and their applications in physics.
  Proceedings, 1994}} 117--136,
\href{http://arxiv.org/abs/hep-th/9408041}{{\ttfamily arXiv:hep-th/9408041
  [hep-th]}}.

\bibitem{Volkov:1992uv}
A.~{\relax Yu}. Volkov, ``{Quantum Volterra model},''
{\em Phys. Lett.} {\bfseries A167} (1992) 345--355.

\bibitem{Bazhanov:2010kz}
V.~V. Bazhanov and S.~M. Sergeev, ``{A Master solution of the quantum
  Yang-Baxter equation and classical discrete integrable equations},''
  \href{http://dx.doi.org/10.4310/ATMP.2012.v16.n1.a3}{{\em Adv. Theor. Math.
  Phys.} {\bfseries 16} no.~1, (2012) 65--95},
\href{http://arxiv.org/abs/1006.0651}{{\ttfamily arXiv:1006.0651 [math-ph]}}.

\bibitem{Spiridonov:2010em}
V.~P. Spiridonov, ``{Elliptic beta integrals and solvable models of statistical
  mechanics},'' {\em Contemp. Math.} {\bfseries 563} (2012) 181--211,
\href{http://arxiv.org/abs/1011.3798}{{\ttfamily arXiv:1011.3798 [hep-th]}}.

\bibitem{Yamazaki:2012cp}
M.~Yamazaki, ``{Quivers, YBE and 3-manifolds},''
  \href{http://dx.doi.org/10.1007/JHEP05(2012)147}{{\em JHEP} {\bfseries 05}
  (2012) 147},
\href{http://arxiv.org/abs/1203.5784}{{\ttfamily arXiv:1203.5784 [hep-th]}}.

\bibitem{Yamazaki:2013nra}
M.~Yamazaki, ``{New Integrable Models from the Gauge/YBE Correspondence},''
  \href{http://dx.doi.org/10.1007/s10955-013-0884-8}{{\em J. Statist. Phys.}
  {\bfseries 154} (2014) 895},
\href{http://arxiv.org/abs/1307.1128}{{\ttfamily arXiv:1307.1128 [hep-th]}}.

\bibitem{Yagi:2015lha}
J.~Yagi, ``{Quiver gauge theories and integrable lattice models},''
  \href{http://dx.doi.org/10.1007/JHEP10(2015)065}{{\em JHEP} {\bfseries 10}
  (2015) 065},
\href{http://arxiv.org/abs/1504.04055}{{\ttfamily arXiv:1504.04055 [hep-th]}}.

\bibitem{Yamazaki:2015voa}
M.~Yamazaki and W.~Yan, ``{Integrability from 2d ${\mathcal{N}}=(2,2)$
  dualities},'' \href{http://dx.doi.org/10.1088/1751-8113/48/39/394001}{{\em J.
  Phys.} {\bfseries A48} (2015) 394001},
\href{http://arxiv.org/abs/1504.05540}{{\ttfamily arXiv:1504.05540 [hep-th]}}.

\bibitem{Gahramanov:2015cva}
I.~Gahramanov and V.~P. Spiridonov, ``{The star-triangle relation and 3d
  superconformal indices},''
  \href{http://dx.doi.org/10.1007/JHEP08(2015)040}{{\em JHEP} {\bfseries 08}
  (2015) 040},
\href{http://arxiv.org/abs/1505.00765}{{\ttfamily arXiv:1505.00765 [hep-th]}}.

\bibitem{Kels:2015bda}
A.~P. Kels, ``{New solutions of the star - triangle relation with discrete and
  continuous spin variables},''
  \href{http://dx.doi.org/10.1088/1751-8113/48/43/435201}{{\em J. Phys.}
  {\bfseries A48} no.~43, (2015) 435201},
\href{http://arxiv.org/abs/1504.07074}{{\ttfamily arXiv:1504.07074 [math-ph]}}.

\bibitem{Maruyoshi:2016caf}
K.~Maruyoshi and J.~Yagi, ``{Surface defects as transfer matrices},''
  \href{http://dx.doi.org/10.1093/ptep/ptw151}{{\em PTEP} {\bfseries 2016}
  no.~11, (2016) 113B01},
\href{http://arxiv.org/abs/1606.01041}{{\ttfamily arXiv:1606.01041 [hep-th]}}.

\bibitem{Gahramanov:2016ilb}
I.~Gahramanov and A.~P. Kels, ``{The star-triangle relation, lens partition
  function, and hypergeometric sum/integrals},''
  \href{http://dx.doi.org/10.1007/JHEP02(2017)040}{{\em JHEP} {\bfseries 02}
  (2017) 040},
\href{http://arxiv.org/abs/1610.09229}{{\ttfamily arXiv:1610.09229 [math-ph]}}.

\bibitem{Yamazaki:2016wnu}
M.~Yamazaki, ``{Cluster-Enriched Yang-Baxter Equation from SUSY Gauge
  Theories},''
\href{http://arxiv.org/abs/1611.07522}{{\ttfamily arXiv:1611.07522 [hep-th]}}.

\bibitem{Yagi:2016oum}
J.~Yagi, ``{Branes and integrable lattice models},''
  \href{http://dx.doi.org/10.1142/S0217732317300038}{{\em Mod. Phys. Lett.}
  {\bfseries A32} no.~03, (2016) 1730003},
\href{http://arxiv.org/abs/1610.05584}{{\ttfamily arXiv:1610.05584 [hep-th]}}.

\bibitem{Yagi:2017hmj}
J.~Yagi, ``{Surface defects and elliptic quantum groups},''
  \href{http://dx.doi.org/10.1007/JHEP06(2017)013}{{\em JHEP} {\bfseries 06}
  (2017) 013},
\href{http://arxiv.org/abs/1701.05562}{{\ttfamily arXiv:1701.05562 [hep-th]}}.

\bibitem{Kels:2017toi}
A.~P. Kels and M.~Yamazaki, ``{Elliptic hypergeometric sum/integral
  transformations and supersymmetric lens index},''
\href{http://arxiv.org/abs/1704.03159}{{\ttfamily arXiv:1704.03159 [math-ph]}}.

\bibitem{Kels:2017fyt}
A.~P. Kels, ``{Exactly solved models on planar graphs with vertices in
  $\mathbb{Z}^3$},'' \href{http://dx.doi.org/10.1088/1751-8121/aa8f68}{{\em J.
  Phys.} {\bfseries A50} no.~49, (2017) 495202},
\href{http://arxiv.org/abs/1705.06528}{{\ttfamily arXiv:1705.06528 [math-ph]}}.

\bibitem{Nekrasov:2009rc}
N.~A. Nekrasov and S.~L. Shatashvili, ``{Quantization of Integrable Systems and
  Four Dimensional Gauge Theories },'' {\em {Proceedings, 16th International
  Congress on Mathematical Physics (ICMP09) }} (2009) 265--289,
\href{http://arxiv.org/abs/0908.4052}{{\ttfamily arXiv:0908.4052 [hep-th]}}.

\bibitem{Nekrasov:2009uh}
N.~A. Nekrasov and S.~L. Shatashvili, ``{Supersymmetric vacua and Bethe
  ansatz},'' \href{http://dx.doi.org/10.1016/j.nuclphysbps.2009.07.047}{{\em
  Nucl. Phys. Proc. Suppl.} {\bfseries 192-193} (2009) 91--112},
\href{http://arxiv.org/abs/0901.4744}{{\ttfamily arXiv:0901.4744 [hep-th]}}.

\bibitem{Bazhanov:2016ajm}
V.~V. Bazhanov, A.~P. Kels, and S.~M. Sergeev, ``{Quasi-classical expansion of
  the star-triangle relation and integrable systems on quad-graphs},'' {\em J.
  Phys.} {\bfseries A49} (2016) to appear,
\href{http://arxiv.org/abs/1602.07076}{{\ttfamily arXiv:1602.07076}}.

\bibitem{perk2006yang}
J.~H. Perk and H.~Au-Yang, ``{Yang-baxter equations },'' {\em Encyclopedia of
  Mathematical Physics, Vol. 5, (Elsevier Science, Oxford, 2006), pp. 465-473,}
  , \href{http://arxiv.org/abs/math-ph/0606053}{{\ttfamily math-ph/0606053}}.

\bibitem{Bellon:1992sf}
M.~P. Bellon, J.~M. Maillard, and C.~Viallet, ``{On the symmetries of
  integrability},''
\href{http://dx.doi.org/10.1142/S021797929200092X}{{\em Int. J. Mod. Phys.}
  {\bfseries B6} (1992) 1881--1904}.

\bibitem{deguchi2003statistical}
T.~Deguchi, ``Introduction to solvable lattice models in statistical and
  mathematical physics,'' {\em Classical and Quantum Nonlinear Integrable
  Systems: Theory and Application} (2003) 107.

\bibitem{Saleur:1990uz}
H.~Saleur and J.~B. Zuber, ``{Integrable lattice models and quantum groups},''
  in {\em {Spring School on String Theory and Quantum Gravity Trieste, Italy,
  23 April -1 May }}, pp.~0001--54.
\newblock
1990.
\newblock

\bibitem{Jafarzade:2017fsc}
S.~Jafarzade and Z.~Nazari, ``{A New Integrable Ising-type Model from 2d
  $\mathcal{N}$=(2,2) Dualities $(2017)$},''
\href{http://arxiv.org/abs/1709.00070}{{\ttfamily arXiv:1709.00070 [hep-th]}}.

\bibitem{Derkachov:2010zz}
S.~E. Derkachov and A.~N. Manashov, ``{General solution of the Yang-Baxter
  equation with symmetry group SL(n,C)},''
  \href{http://dx.doi.org/10.1090/S1061-0022-2010-01106-3}{{\em St. Petersburg
  Math. J.} {\bfseries 21} (2010) 513--577}.
[Alg. Anal.21N4,1(2009)].

\bibitem{Derkachov:2012iv}
S.~E. Derkachov and V.~P. Spiridonov, ``{Yang-Baxter equation, parameter
  permutations, and the elliptic beta integral},''
  \href{http://dx.doi.org/10.1070/RM2013v068n06ABEH004869}{{\em Russ. Math.
  Surveys} {\bfseries 68} (2013) 1027--1072},
\href{http://arxiv.org/abs/1205.3520}{{\ttfamily arXiv:1205.3520 [math-ph]}}.

\bibitem{Chicherin:2014dya}
D.~Chicherin, S.~E. Derkachov, and V.~P. Spiridonov, ``{New elliptic solutions
  of the Yang-Baxter equation},''
\href{http://arxiv.org/abs/1412.3383}{{\ttfamily arXiv:1412.3383 [math-ph]}}.

\bibitem{baxter1980hard}
R.~J. Baxter, ``Hard hexagons: exact solution,''
  \href{http://dx.doi.org/10.1088/0305-4470/13/3/007}{{\em Journal of Physics
  A: Mathematical and General} {\bfseries 13} no.~3, (1980) L61}.

\bibitem{Baxter:1972wg}
R.~J. Baxter, ``{Eight vertex model in lattice statistics and one-dimensional
  anisotropic Heisenberg chain. 1. Some fundamental eigenvectors},''
\href{http://dx.doi.org/10.1016/0003-4916(73)90439-9}{{\em Annals Phys.}
  {\bfseries 76} (1973) 1--24}.

\bibitem{Baxter:1972wf}
R.~J. Baxter, ``{Eight vertex model in lattice statistics and one-dimensional
  anisotropic Heisenberg chain. 2. Equivalence to a generalized ice-type
  lattice model},''
\href{http://dx.doi.org/10.1016/0003-4916(73)90440-5}{{\em Annals Phys.}
  {\bfseries 76} (1973) 25--47}.

\bibitem{Baxter:1972wh}
R.~J. Baxter, ``{Eight vertex model in lattice statistics and one-dimensional
  anisotropic Heisenberg chain. 1. Eigenvectors of the transfer matrix and
  Hamiltonian},''
\href{http://dx.doi.org/10.1016/0003-4916(73)90441-7}{{\em Annals Phys.}
  {\bfseries 76} (1973) 48--71}.

\bibitem{Pearce:1988en}
P.~A. Pearce and K.~A. Seaton, ``{A solvable hierarchy of cyclic solid-on-solid
  lattice models},''
\href{http://dx.doi.org/10.1103/PhysRevLett.60.1347}{{\em Phys. Rev. Lett.}
  {\bfseries 60} (1988) 1347--1350}.

\bibitem{Pearce:1989ek}
P.~A. Pearce and K.~A. Seaton, ``{Exact solution of cyclic solid-on-solid
  lattice models},''
\href{http://dx.doi.org/10.1016/0003-4916(89)90003-1}{{\em Annals Phys.}
  {\bfseries 193} (1989) 326--366}.

\bibitem{Andrews:1984af}
G.~E. Andrews, R.~J. Baxter, and P.~J. Forrester, ``{Eight vertex SOS model and
  generalized Rogers-Ramanujan type identities},''
\href{http://dx.doi.org/10.1007/BF01014383}{{\em J. Statist. Phys.} {\bfseries
  35} (1984) 193--266}.

\bibitem{Lieb:1967zz}
E.~H. Lieb, ``{Residual Entropy of Square Ice},''
\href{http://dx.doi.org/10.1103/PhysRev.162.162}{{\em Phys. Rev.} {\bfseries
  162} (1967) 162--172}.

\bibitem{sutherland1967exact}
B.~Sutherland, ``Exact solution of a two-dimensional model for hydrogen-bonded
  crystals,'' {\em Physical Review Letters} {\bfseries 19} no.~3, (1967) 103.

\bibitem{Izergin:1980pe}
A.~G. Izergin and V.~E. Korepin, ``{The inverse scattering method approach to
  the quantum Shabat-Mikhailov model},''
\href{http://dx.doi.org/10.1007/BF01208496}{{\em Commun. Math. Phys.}
  {\bfseries 79} (1981) 303}.

\bibitem{Perk:1986nr}
J.~H.~H. Perk and F.~Y. Wu, ``{Graphical Approach to the Nonintersecting String
  Model: Star Triangle Equation, Inversion Relation and Exact Solution},''
\href{http://dx.doi.org/10.1016/0378-4371(86)90175-5}{{\em Physica} {\bfseries
  A138} (1986) 100--124}.

\bibitem{Baxter:1982xp}
R.~J. Baxter, ``{The inversion relation method for some two-dimensional exactly
  solved models in lattice statistics},''
\href{http://dx.doi.org/10.1007/BF01011621}{{\em J. Statist. Phys.} {\bfseries
  28} (1982) 1--41}.

\bibitem{Intriligator:1995au}
K.~A. Intriligator and N.~Seiberg, ``{Lectures on supersymmetric gauge theories
  and electric-magnetic duality},''
  \href{http://dx.doi.org/10.1016/0920-5632(95)00626-5}{{\em Nucl. Phys. Proc.
  Suppl.} {\bfseries 45BC} (1996) 1--28},
\href{http://arxiv.org/abs/hep-th/9509066}{{\ttfamily arXiv:hep-th/9509066
  [hep-th]}}.

\bibitem{Strassler:2001ue}
M.~J. Strassler, ``{Erice Lectures on Confinement and Duality},''
  \href{http://dx.doi.org/10.1142/9789812796653_0004}{{\em Subnucl. Ser.}
  {\bfseries 40} (2003) 154--193}.
[ICTP Lect. Notes Ser.7,105(2002)].

\bibitem{Strassler:2005qs}
M.~J. Strassler, ``{The Duality cascade},''
  \href{http://dx.doi.org/10.1142/9789812775108_0005}{{\em {Progress in string
  theory. Proceedings, Summer School, TASI 2003, Boulder, USA, June 2-27,
  2003}} (2005) 419--510},
\href{http://arxiv.org/abs/hep-th/0505153}{{\ttfamily arXiv:hep-th/0505153
  [hep-th]}}.

\bibitem{Terning:2006bq}
J.~Terning,
  \href{http://dx.doi.org/10.1093/acprof:oso/9780198567639.001.0001}{{\em
  {Modern supersymmetry: Dynamics and duality}}}.
\newblock
2006.
\newblock

\bibitem{Seiberg:1994pq}
N.~Seiberg, ``{Electric - magnetic duality in supersymmetric nonAbelian gauge
  theories},'' \href{http://dx.doi.org/10.1016/0550-3213(94)00023-8}{{\em Nucl.
  Phys.} {\bfseries B435} (1995) 129--146},
\href{http://arxiv.org/abs/hep-th/9411149}{{\ttfamily arXiv:hep-th/9411149
  [hep-th]}}.

\bibitem{Berenstein:2002fi}
D.~Berenstein and M.~R. Douglas, ``{Seiberg duality for quiver gauge
  theories},''
\href{http://arxiv.org/abs/hep-th/0207027}{{\ttfamily arXiv:hep-th/0207027
  [hep-th]}}.

\bibitem{Yamazaki:2008bt}
M.~Yamazaki, ``{Brane Tilings and Their Applications},''
  \href{http://dx.doi.org/10.1002/prop.200810536}{{\em Fortsch. Phys.}
  {\bfseries 56} (2008) 555--686},
\href{http://arxiv.org/abs/0803.4474}{{\ttfamily arXiv:0803.4474 [hep-th]}}.

\bibitem{Hanany:2005ss}
A.~Hanany and D.~Vegh, ``{Quivers, tilings, branes and rhombi},''
  \href{http://dx.doi.org/10.1088/1126-6708/2007/10/029}{{\em JHEP} {\bfseries
  10} (2007) 029},
\href{http://arxiv.org/abs/hep-th/0511063}{{\ttfamily arXiv:hep-th/0511063
  [hep-th]}}.

\bibitem{Gahramanov:2015tta}
I.~Gahramanov, ``{Mathematical structures behind supersymmetric dualities},''
  \href{http://dx.doi.org/10.5817/AM2015-5-273}{{\em Archivum Math.} {\bfseries
  51} (2015) 273--286},
\href{http://arxiv.org/abs/1505.05656}{{\ttfamily arXiv:1505.05656 [math-ph]}}.

\bibitem{Gahramanov:gka}
I.~B. Gahramanov and G.~S. Vartanov, ``{Superconformal indices and partition
  functions for supersymmetric field theories},''
  \href{http://dx.doi.org/10.1142/9789814449243_0076}{{\em {XVIIth Intern.
  Cong. Math. Phys. 695-703}} (2013) },
\href{http://arxiv.org/abs/1310.8507}{{\ttfamily arXiv:1310.8507 [hep-th]}}.

\bibitem{Yamazaki:2013fva}
M.~Yamazaki, ``{Four-dimensional superconformal index reloaded},''
  \href{http://dx.doi.org/10.1007/s11232-013-0012-6}{{\em Theor. Math. Phys.}
  {\bfseries 174} (2013) 154--166}.
[Teor. Mat. Fiz.174,177(2013)].

\bibitem{Aharony:2013dha}
O.~Aharony, S.~S. Razamat, N.~Seiberg, and B.~Willett, ``{3d dualities from 4d
  dualities},'' \href{http://dx.doi.org/10.1007/JHEP07(2013)149}{{\em JHEP}
  {\bfseries 07} (2013) 149},
\href{http://arxiv.org/abs/1305.3924}{{\ttfamily arXiv:1305.3924 [hep-th]}}.

\bibitem{Aharony:2017adm}
O.~Aharony, S.~S. Razamat, and B.~Willett, ``{From 3d duality to 2d duality},''
  \href{http://dx.doi.org/10.1007/JHEP11(2017)090}{{\em JHEP} {\bfseries 11}
  (2017) 090},
\href{http://arxiv.org/abs/1710.00926}{{\ttfamily arXiv:1710.00926 [hep-th]}}.

\bibitem{Spiridonov:2014cxa}
V.~P. Spiridonov and G.~S. Vartanov, ``{Vanishing superconformal indices and
  the chiral symmetry breaking},''
  \href{http://dx.doi.org/10.1007/JHEP06(2014)062}{{\em JHEP} {\bfseries 06}
  (2014) 062},
\href{http://arxiv.org/abs/1402.2312}{{\ttfamily arXiv:1402.2312 [hep-th]}}.

\bibitem{Elitzur:1997fh}
S.~Elitzur, A.~Giveon, and D.~Kutasov, ``{Branes and N=1 duality in string
  theory},'' \href{http://dx.doi.org/10.1016/S0370-2693(97)00375-4}{{\em Phys.
  Lett.} {\bfseries B400} (1997) 269--274},
\href{http://arxiv.org/abs/hep-th/9702014}{{\ttfamily arXiv:hep-th/9702014
  [hep-th]}}.

\bibitem{Elitzur:1997hc}
S.~Elitzur, A.~Giveon, D.~Kutasov, E.~Rabinovici, and A.~Schwimmer, ``{Brane
  dynamics and N=1 supersymmetric gauge theory},''
  \href{http://dx.doi.org/10.1016/S0550-3213(97)00446-X}{{\em Nucl. Phys.}
  {\bfseries B505} (1997) 202--250},
\href{http://arxiv.org/abs/hep-th/9704104}{{\ttfamily arXiv:hep-th/9704104
  [hep-th]}}.

\bibitem{Giveon:1998sr}
A.~Giveon and D.~Kutasov, ``{Brane dynamics and gauge theory},''
  \href{http://dx.doi.org/10.1103/RevModPhys.71.983}{{\em Rev. Mod. Phys.}
  {\bfseries 71} (1999) 983--1084},
\href{http://arxiv.org/abs/hep-th/9802067}{{\ttfamily arXiv:hep-th/9802067
  [hep-th]}}.

\bibitem{Bazhanov:2013bh}
V.~V. Bazhanov, A.~P. Kels, and S.~M. Sergeev, ``{Comment on star-star
  relations in statistical mechanics and elliptic gamma-function identities},''
  \href{http://dx.doi.org/10.1088/1751-8113/46/15/152001}{{\em J. Phys.}
  {\bfseries A46} (2013) 152001},
\href{http://arxiv.org/abs/1301.5775}{{\ttfamily arXiv:1301.5775 [math-ph]}}.

\bibitem{Bazhanov:2011mz}
V.~V. Bazhanov and S.~M. Sergeev, ``{Elliptic gamma-function and multi-spin
  solutions of the Yang-Baxter equation},''
  \href{http://dx.doi.org/10.1016/j.nuclphysb.2011.10.032}{{\em Nucl. Phys.}
  {\bfseries B856} (2012) 475--496},
\href{http://arxiv.org/abs/1106.5874}{{\ttfamily arXiv:1106.5874 [math-ph]}}.

\bibitem{Gahramanov:2017idz}
I.~Gahramanov and S.~Jafarzade, ``{Comments on the multi-spin solution to the
  Yang-Baxter equation and basic hypergeometric sum/integral identity
  $(2017)$},''
\href{http://arxiv.org/abs/1710.09106}{{\ttfamily arXiv:1710.09106 [math-ph]}}.

\bibitem{Dolan:2008qi}
F.~A. Dolan and H.~Osborn, ``{Applications of the Superconformal Index for
  Protected Operators and q-Hypergeometric Identities to N=1 Dual Theories},''
  \href{http://dx.doi.org/10.1016/j.nuclphysb.2009.01.028}{{\em Nucl. Phys.}
  {\bfseries B818} (2009) 137--178},
\href{http://arxiv.org/abs/0801.4947}{{\ttfamily arXiv:0801.4947 [hep-th]}}.

\bibitem{Spiridonovbeta}
V.~P. Spiridonov, ``On the elliptic beta function,'' {\em Russian Mathematical
  Surveys} {\bfseries 56} no.~1, (2001) 185.
  \url{http://stacks.iop.org/0036-0279/56/i=1/a=L21}.

\bibitem{Kapustin:2009kz}
A.~Kapustin, B.~Willett, and I.~Yaakov, ``{Exact Results for Wilson Loops in
  Superconformal Chern-Simons Theories with Matter},''
  \href{http://dx.doi.org/10.1007/JHEP03(2010)089}{{\em JHEP} {\bfseries 03}
  (2010) 089},
\href{http://arxiv.org/abs/0909.4559}{{\ttfamily arXiv:0909.4559 [hep-th]}}.

\bibitem{Gahramanov:2013xsa}
I.~Gahramanov and G.~Vartanov, ``{Extended global symmetries for 4D $N$ = 1
  SQCD theories},''
  \href{http://dx.doi.org/10.1088/1751-8113/46/28/285403}{{\em J. Phys.}
  {\bfseries A46} (2013) 285403},
\href{http://arxiv.org/abs/1303.1443}{{\ttfamily arXiv:1303.1443 [hep-th]}}.

\bibitem{Gahramanov:2013rda}
I.~Gahramanov and H.~Rosengren, ``{A new pentagon identity for the tetrahedron
  index},'' \href{http://dx.doi.org/10.1007/JHEP11(2013)128}{{\em JHEP}
  {\bfseries 11} (2013) 128},
\href{http://arxiv.org/abs/1309.2195}{{\ttfamily arXiv:1309.2195 [hep-th]}}.

\bibitem{Gahramanov:2014ona}
I.~Gahramanov and H.~Rosengren, ``{Integral pentagon relations for 3d
  superconformal indices},''
  \href{http://dx.doi.org/10.1090/pspum/093/01569}{{\em {String-Math 2014:
  Proceedings, Alberta, Canada, June 9-13}} (2014) },
\href{http://arxiv.org/abs/1412.2926}{{\ttfamily arXiv:1412.2926 [hep-th]}}.

\bibitem{Gahramanov:2016wxi}
I.~Gahramanov and H.~Rosengren, ``{Basic hypergeometry of supersymmetric
  dualities},'' \href{http://dx.doi.org/10.1016/j.nuclphysb.2016.10.004}{{\em
  Nucl. Phys.} {\bfseries B913} (2016) 747--768},
\href{http://arxiv.org/abs/1606.08185}{{\ttfamily arXiv:1606.08185 [hep-th]}}.

\bibitem{Rosengren:2016mnw}
H.~Rosengren, ``{Rahman's biorthogonal functions and superconformal indices},''
\href{http://arxiv.org/abs/1612.05051}{{\ttfamily arXiv:1612.05051 [math.CA]}}.

\bibitem{Imamura:2013nra}
Y.~Imamura and D.~Yokoyama, ``{$\mathcal{N} = 2$ supersymmetric theories on
  squashed three-sphere},''
\href{http://dx.doi.org/10.1142/S2010194513009665}{{\em Int. J. Mod. Phys.
  Conf. Ser.} {\bfseries 21} (2013) 171--172}.

\bibitem{Hosomichi:2014hja}
K.~Hosomichi, ``{A Review on SUSY Gauge Theories on $\mathbf{S^3}$},''
  \href{http://dx.doi.org/10.1007/978-3-319-18769-3_10}{{\em New Dualities of
  Supersymmetric Gauge Theories} (2016) 307--338},
  \href{http://arxiv.org/abs/1412.7128}{{\ttfamily arXiv:1412.7128 [hep-th]}}.
\url{https://inspirehep.net/record/1335341/files/arXiv:1412.7128.pdf}.

\bibitem{Nian:2013qwa}
J.~Nian, ``{Localization of Supersymmetric Chern-Simons-Matter Theory on a
  Squashed $S^3$ with $SU(2)\times U(1)$ Isometry},''
  \href{http://dx.doi.org/10.1007/JHEP07(2014)126}{{\em JHEP} {\bfseries 07}
  (2014) 126},
\href{http://arxiv.org/abs/1309.3266}{{\ttfamily arXiv:1309.3266 [hep-th]}}.

\bibitem{Hama:2011ea}
N.~Hama, K.~Hosomichi, and S.~Lee, ``{SUSY Gauge Theories on Squashed
  Three-Spheres},'' \href{http://dx.doi.org/10.1007/JHEP05(2011)014}{{\em JHEP}
  {\bfseries 05} (2011) 014},
\href{http://arxiv.org/abs/1102.4716}{{\ttfamily arXiv:1102.4716 [hep-th]}}.

\bibitem{Imamura:2011wg}
Y.~Imamura and D.~Yokoyama, ``{N=2 supersymmetric theories on squashed
  three-sphere},'' \href{http://dx.doi.org/10.1103/PhysRevD.85.025015}{{\em
  Phys. Rev.} {\bfseries D85} (2012) 025015},
\href{http://arxiv.org/abs/1109.4734}{{\ttfamily arXiv:1109.4734 [hep-th]}}.

\bibitem{Teschner:2012em}
J.~Teschner and G.~Vartanov, ``{6j symbols for the modular double, quantum
  hyperbolic geometry, and supersymmetric gauge theories},''
  \href{http://dx.doi.org/10.1007/s11005-014-0684-3}{{\em Lett. Math. Phys.}
  {\bfseries 104} (2014) 527--551},
\href{http://arxiv.org/abs/1202.4698}{{\ttfamily arXiv:1202.4698 [hep-th]}}.

\bibitem{Dolan:2011rp}
F.~A.~H. Dolan, V.~P. Spiridonov, and G.~S. Vartanov, ``{From 4d superconformal
  indices to 3d partition functions},''
  \href{http://dx.doi.org/10.1016/j.physletb.2011.09.007}{{\em Phys. Lett.}
  {\bfseries B704} (2011) 234--241},
\href{http://arxiv.org/abs/1104.1787}{{\ttfamily arXiv:1104.1787 [hep-th]}}.

\bibitem{Bazhanov:2007mh}
V.~V. Bazhanov, V.~V. Mangazeev, and S.~M. Sergeev, ``{Faddeev-Volkov solution
  of the Yang-Baxter equation and discrete conformal symmetry},''
  \href{http://dx.doi.org/10.1016/j.nuclphysb.2007.05.013}{{\em Nucl. Phys.}
  {\bfseries B784} (2007) 234--258},
\href{http://arxiv.org/abs/hep-th/0703041}{{\ttfamily arXiv:hep-th/0703041
  [hep-th]}}.

\bibitem{Bazhanov:2007vg}
V.~V. Bazhanov, V.~V. Mangazeev, and S.~M. Sergeev, ``{Exact solution of the
  Faddeev-Volkov model},''
  \href{http://dx.doi.org/10.1016/j.physleta.2007.10.053}{{\em Phys. Lett.}
  {\bfseries A372} (2008) 1547--1550},
\href{http://arxiv.org/abs/0706.3077}{{\ttfamily arXiv:0706.3077
  [cond-mat.stat-mech]}}.

\bibitem{Bultthesis}
F.~J. van~de Bult, {\em Hyperbolic hypergeometric functions}.
\newblock PhD thesis, University of Amsterdam, 2007.

\bibitem{Kels:2013ola}
A.~P. Kels, ``{A new solution of the star-triangle relation},''
  \href{http://dx.doi.org/10.1088/1751-8113/47/5/055203}{{\em J. Phys.}
  {\bfseries A47} (2014) 055203},
\href{http://arxiv.org/abs/1302.3025}{{\ttfamily arXiv:1302.3025 [math-ph]}}.

\bibitem{Benini:2013xpa}
F.~Benini, R.~Eager, K.~Hori, and Y.~Tachikawa, ``{Elliptic Genera of 2d
  ${\mathcal{N}}$ = 2 Gauge Theories},''
  \href{http://dx.doi.org/10.1007/s00220-014-2210-y}{{\em Commun. Math. Phys.}
  {\bfseries 333} no.~3, (2015) 1241--1286},
\href{http://arxiv.org/abs/1308.4896}{{\ttfamily arXiv:1308.4896 [hep-th]}}.

\bibitem{Benini:2013nda}
F.~Benini, R.~Eager, K.~Hori, and Y.~Tachikawa, ``{Elliptic genera of
  two-dimensional N=2 gauge theories with rank-one gauge groups},''
  \href{http://dx.doi.org/10.1007/s11005-013-0673-y}{{\em Lett. Math. Phys.}
  {\bfseries 104} (2014) 465--493},
\href{http://arxiv.org/abs/1305.0533}{{\ttfamily arXiv:1305.0533 [hep-th]}}.

\bibitem{Gadde:2013ftv}
A.~Gadde and S.~Gukov, ``{2d Index and Surface operators},''
  \href{http://dx.doi.org/10.1007/JHEP03(2014)080}{{\em JHEP} {\bfseries 03}
  (2014) 080},
\href{http://arxiv.org/abs/1305.0266}{{\ttfamily arXiv:1305.0266 [hep-th]}}.

\bibitem{GahJaf2}
I.~Gahramanov, S.~Jafarzade, A.~Kels, and G.~Mogol {\em to appear.} .

\bibitem{Kuniba:2015azm}
A.~Kuniba, M.~Okado, and S.~Sergeev, ``{Tetrahedron equation and generalized
  quantum groups},''
  \href{http://dx.doi.org/10.1088/1751-8113/48/30/304001}{{\em J. Phys.}
  {\bfseries A48} no.~30, (2015) 304001},
\href{http://arxiv.org/abs/1503.08536}{{\ttfamily arXiv:1503.08536 [math.QA]}}.

\bibitem{Bazhanov:2005as}
V.~V. Bazhanov and S.~M. Sergeev, ``{Zamolodchikov's tetrahedron equation and
  hidden structure of quantum groups},''
  \href{http://dx.doi.org/10.1088/0305-4470/39/13/009}{{\em J. Phys.}
  {\bfseries A39} (2006) 3295--3310},
\href{http://arxiv.org/abs/hep-th/0509181}{{\ttfamily arXiv:hep-th/0509181
  [hep-th]}}.

\bibitem{Stroganov:1997br}
{\relax Yu}.~G. Stroganov, ``{Tetrahedron equation and spin integrable models
  on a cubic lattice},'' \href{http://dx.doi.org/10.1007/BF02630441}{{\em
  Theor. Math. Phys.} {\bfseries 110} (1997) 141--167}.
[Teor. Mat. Fiz.110,179(1997)].

\bibitem{Bazhanov:1981zm}
V.~V. Bazhanov and {\relax Yu}.~G. Stroganov, ``{On Commutativity Conditions
  for Transfer Matrices on Multidimensional Lattice},''
  \href{http://dx.doi.org/10.1007/BF01027789}{{\em Theor. Math. Phys.}
  {\bfseries 52} (1982) 685--691}.
[Teor. Mat. Fiz.52,105(1982)].

\bibitem{frenkel1991}
I.~Frenkel and G.~Moore, ``Simplex equations and their solutions,'' {\em Comm.
  Math. Phys.} {\bfseries 138} no.~2, (1991) 259--271.

\bibitem{zamolodchikov1980tetrahedra}
A.~Zamolodchikov, ``Tetrahedra equations and integrable systems in
  three-dimensional space,'' {\em Soviet Journal of Experimental and
  Theoretical Physics} {\bfseries 52} (1980) 325.

\bibitem{fomin2002cluster}
S.~Fomin and A.~Zelevinsky, ``Cluster algebras i: foundations,'' {\em Journal
  of the American Mathematical Society} {\bfseries 15} no.~2, (2002) 497--529,
  \href{http://arxiv.org/abs/math/0104151}{{\ttfamily math/0104151}}.

\bibitem{Ruijsenaars:1997:FOA}
S.~N.~M. Ruijsenaars, ``First order analytic difference equations and
  integrable quantum systems,'' \href{http://dx.doi.org/10.1063/1.531809}{{\em
  J. Math. Phys.} {\bfseries 38} no.~2, (1997) 1069--1146}.

\bibitem{Felder200044}
G.~Felder and A.~Varchenko, ``{The Elliptic Gamma Function and $SL(3, Z) \times
  Z_3$},''
  \href{http://dx.doi.org/http://dx.doi.org/10.1006/aima.2000.1951}{{\em Adv.
  Math.} {\bfseries 156} no.~1, (2000) 44 -- 76},
  \href{http://arxiv.org/abs/math/9907061}{{\ttfamily arXiv:math/9907061}}.

\bibitem{friedman2004shintani}
E.~Friedman and S.~Ruijsenaars, ``Shintani--barnes zeta and gamma functions,''
  {\em Advances in Mathematics} {\bfseries 187} no.~2, (2004) 362--395.

\end{thebibliography}\endgroup

\end{document}